\documentclass[useAMS,usenatbib]{mn2e}
\usepackage{graphicx,epsfig}

\title[Gas stripping of NGC~2276]{Gas stripping in galaxy groups --
  the case of the starburst spiral NGC~2276} \author[J. Rasmussen, T.
J. Ponman and J. S. Mulchaey]{Jesper Rasmussen,$^{1}$\thanks{E-mail:
    jesper@star.sr.bham.ac.uk} Trevor J. Ponman$^{1}$
  and John S. Mulchaey$^{2}$\\
  $^{1}$School of Physics and Astronomy, University of Birmingham,
  Edgbaston,
  Birmingham B15 2TT\\
  $^{2}$Observatories of the Carnegie Institution, 813 Santa Barbara
  Street, Pasadena, California, USA}
\begin{document}

\date{Accepted ---. Received ---; in original form ---}

\pagerange{\pageref{firstpage}--\pageref{lastpage}} \pubyear{2006}

\maketitle

\label{firstpage}

\begin{abstract} 
  Ram pressure stripping of galactic gas is generally assumed to be
  inefficient in galaxy groups due to the relatively low density of
  the intragroup medium and the small velocity dispersions of groups.
  To test this assumption, we obtained {\em Chandra} X-ray data of the
  starbursting spiral NGC~2276 in the NGC~2300 group of galaxies, a
  candidate for a strong galaxy interaction with hot intragroup gas.
  The data reveal a shock-like feature along the western edge of the
  galaxy and a low--surface-brightness tail extending to the east,
  similar to the morphology seen in other wavebands.  Spatially
  resolved spectroscopy shows that the data are consistent with
  intragroup gas being pressurized at the leading western edge of
  NGC~2276 due to the galaxy moving supersonically through the
  intragroup medium at a velocity $\sim 850$~km~s$^{-1}$.  Detailed
  modelling of the gravitational potential of NGC~2276 shows that the
  resulting ram-pressure could significantly affect the morphology of
  the outer gas disc but is probably insufficient to strip large
  amounts of cold gas from the disc. We estimate the mass loss rates
  due to turbulent viscous stripping and starburst outflows being
  swept back by ram pressure, showing that both mechanisms could
  plausibly explain the presence of the X-ray tail. Comparison to
  existing H{\sc i} measurements shows that most of the gas escaping
  the galaxy is in a hot phase. With a total mass loss rate of $\sim
  5$~M$_\odot$~yr$^{-1}$, the galaxy could be losing its entire
  present H{\sc i} supply within a Gyr. This demonstrates that the
  removal of galactic gas through interactions with a hot intragroup
  medium can occur rapidly enough to transform the morphology of
  galaxies in groups.  Implications of this for galaxy evolution in
  groups and clusters are briefly discussed.
\end{abstract}


\section{Introduction}\label{sec,intro}

There is a long standing argument about which processes are the key
ones in modifying the morphology of galaxies in dense environments,
and establishing the morphology-density relation and the associated
truncation of star formation (see, e.g., \citealt{goto03}).  Candidate
processes for the morphological evolution of disc galaxies into
lenticulars include ram pressure stripping, tidal stripping, galaxy
harassment, strangulation, interaction-induced star formation, and
major and minor mergers. Of these, the first two mechanisms have been
extensively studied using both observational (e.g.\ \citealt{vogt04})
and numerical approaches \citep*{abad99,stev99,quil00,roed05}.  In
particular, recent H{\sc i} and X-ray observations of individual
cluster galaxies have uncovered strong evidence for ongoing
ram-pressure stripping (e.g.\ \citealt{voll04}; \citealt*{wang04};
\citealt{sun05}).

While most studies of gas stripping have focused on galaxies in
clusters, including the statistical approach presented by
\citet{goto05}, there is growing evidence that much of the
modification of galaxy morphologies and truncation of star formation
actually takes place in groups and cluster outskirts rather than in
the dense cluster cores \citep{lewi02,hels03,gonz05,home06}.  This has
been interpreted by many authors as ruling out the possibility that
stripping via interactions with a hot intracluster medium (ICM) plays
a major role in morphological transformations, given that it is
expected to be much less effective in poor systems than in the cores
of rich clusters.  This interpretation remains to be tested by direct
observations, however, and recently, claims have been made for X-ray
(\citealt*{siva04}; see also \citealt{mach05b}) and H{\sc i}
\citep{bure02,kant05} evidence of ram-pressure stripping in groups.
But, as noted by these authors, ram-pressure stripping is not the only
viable explanation for the gas morphology seen in these systems. One
problem with assessing the importance of ram pressure in these cases
is that of obtaining robust constraints on the true 3-D velocity of
the galaxy through the intragroup medium.

In an attempt to investigate the efficiency of gas removal in groups,
and in particular test whether gas stripping through ICM interactions
can in fact be effective in group environments, we obtained {\em
  Chandra} data of the spiral NGC~2276 in the NGC~2300 group of
galaxies. This galaxy displays a disturbed optical morphology and
constitutes another good candidate for a strong interaction with hot
intragroup gas. NGC~2300 itself is the central elliptical in a sparse
group catalogued with just four galaxies \citep{giud99}, this group
being the first in which an X-ray emitting intragroup medium (IGM) was
detected \citep{mulc93}.  The peculiar morphology of NGC~2276 has also
attracted attention, spawning detailed studies of its optical, radio,
and X-ray point source properties \citep{grue93,humm95,davi97,davi04}.
The distributions of optical light, H{\sc i} gas, and radio continuum
emission in this galaxy show a bow-shock--like structure along the
western edge, which may suggest a shocked gas component. If so, the
galaxy must be moving supersonically through the ambient IGM.
Evidence supporting this interpretation is the high star formation
rate of NGC~2276 ($\sim 5$~M$_\odot$~yr$^{-1}$; \citealt{davi97}),
with much of this starburst activity occurring along the western edge,
as expected if an interaction with the IGM has triggered some of the
star formation via ram pressure compression of molecular gas.

Also, similarly to many of the Virgo cluster spirals showing signs of
ICM interactions \citep{caya90}, there is evidence that NGC~2276 is
deficient in H{\sc i} compared to spirals in the field. For an
isolated spiral of the same morphology and optical luminosity, the
relation of \citet{hayn84} would predict an H{\sc i} content $M_{\rm
  HI,pred.}\approx 1.4\times 10^{10}$~M$_\odot$, which is 2.2 times
the observed amount of $M_{\rm HI, obs.} \approx 6.4\times
10^9$~M$_\odot$ (\citealt{youn89}; corrected to the adopted distance
of 36.8~Mpc, see below). This corresponds to an H{\sc i} deficiency,
as defined by \citet{verd01}, of log $M_{\rm HI,pred.} - \mbox{log}
M_{\rm HI,obs.} \approx 0.35$, and suggests that NGC~2276 may already
have lost a significant fraction of its original (i.e.\ prior to
infall into the group) gas content.

To estimate the pressure conditions across the possible shock front
and settle the question of whether a bow shock is really present in
NGC~2276, X-ray data of high spatial resolution remain the only viable
means.  These can enable constraints to be placed on the 3-D velocity
of the galaxy and allow a test of whether the conditions for
ram-pressure stripping are met.  This is the goal of the present
study. To this end, we use {\em Chandra} data to map out the detailed
spectral and spatial features of hot gas in and around NGC~2276 and
NGC~2300. Salient features of the observed group galaxies are listed
in Table~\ref{tab,gals}.

Adopting $H_0=75$ km s$^{-1}$ Mpc$^{-1}$, the distance to NGC~2276 is
36.8 Mpc \citep{tull88}, with 1 arcmin corresponding to 10.7 kpc. All
errors are quoted at 90 per cent confidence unless stated otherwise.

\begin{table}
\begin{center}
  \caption{Properties of observed galaxies. $B$- and $K$-band
    luminosities $L_B$ and $L_K$ calculated from
    (extinction-corrected) apparent magnitudes listed in
    \citet{tull88} and the NASA/IPAC Extragalactic Database (NED),
    respectively.  Stellar masses $M_\ast$ computed using the
    prescription of \citet{mann05}.  Morphologies and recession
    velocities $v_r$ (corrected for Virgocentric infall) from the
    HyperLeda database.}
\label{tab,gals}
 \begin{tabular}{@{}lcc}
  \hline
Galaxy                 &    NGC~2276    & NGC~2300       \\ \hline
RA (J2000)             &   07  27 11.42 &   07 32 20.05  \\
Dec (J2000)            & $+85$ 45 19.0  & $+85$ 42 31.4  \\
Hubble type            & SBc & E3 \\
 $L_B$ (L$_{B,\odot}$) & $4.4\times 10^{10}$  &  $4.2\times 10^{10}$ \\
 $L_K$ (L$_{K,\odot}$) & $7.0\times 10^{10}$  & $2.7\times 10^{11}$ \\
$M_{\ast}$ (M$_\odot$) & $2.7\times 10^{10}$  & $2.1\times 10^{11}$ \\
$v_r$ (km s$^{-1}$)    & 2684 & 2263 \\
\hline
\end{tabular}
\end{center}
\end{table}

\section{Observations and analysis}

\subsection{Data preparation}

NGC~2276 was observed by {\em Chandra} (obs.\ ID 4968) with the
ACIS-S3 chip as aimpoint, for an effective exposure time of 45.8 ks,
and with the CCDs at a temperature of $-120^{\circ}$~C. The nearby
group elliptical NGC~2300, which is the brightest group galaxy, was
located on the S2 chip.  The two other galaxies in the group, NGC~2268
and IC~455, are not covered by this {\em Chandra} pointing.  A fifth
galaxy, UGC~03670, with a distance to NGC~2300 of 16~arcmin ($\sim
170$~kpc), has a redshift concordant with the other group galaxies and
is probably yet another group member. This galaxy is located on the I3
chip; showing only very faint X-ray emission it will not be discussed
further here, where we will concentrate on the emission on the S2 and
S3 chips.

Data were telemetered in Very Faint mode which allows for a superior
suppression of background events, so calibrated event lists were
regenerated and background screened using the {\sc
  acis\_process\_events} tool in {\sc ciao} v3.2. Bad pixels were
screened out using the bad pixel map provided by the pipeline, and
remaining events were grade filtered, excluding {\em ASCA} grades 1,
5, and 7. Periods of high background were identified using $3\sigma$
clipping of full-chip lightcurves, binned in time bins of length
259.8-s and extracted in off-source regions in the 2.5--7~keV band for
back-illuminated chips and in 0.3--12 keV for front-illuminated chips.
These periods were excluded from the data, leaving a cleaned exposure
time of 44.2~ks.  Blank-sky background data from the calibration
database were screened and filtered as for source data.

Point source searches were carried out with {\sc wavdetect} using a
range of scales and detection thresholds. Results were combined,
yielding a total of 112 detected sources, 75 of which are located on
the S2 and S3 chips used in this analysis.  Source extents were
quantified using the $4\sigma$ detection ellipses from {\sc
  wavdetect}, and these regions were masked out in all subsequent
analysis.

The {\em Chandra} astrometry was checked against the optical position
of NGC~2300 as listed in NED, revealing no significant difference.

\subsection{Imaging and spectral analysis}\label{sec,analysis}

Full-resolution (0.5~arcsec~pixel$^{-1}$) adaptively smoothed,
background subtracted images are shown in Fig.~\ref{fig,N2276}. To
produce these, background maps were generated from blank-sky data, and
scaled to match the source count rates in source--free regions on the
relevant chip. They were then smoothed to the same spatial scales as
the source data and subtracted from the latter.  The resulting images
were finally exposure corrected using similarly smoothed, spectrally
weighted exposure maps, with weights derived from spectral fits to the
integrated diffuse emission. We stress that the resulting images
served illustration purposes only and were not used for any
quantitative analyses.
\begin{figure*}
\begin{center}
 \includegraphics[width=135mm]{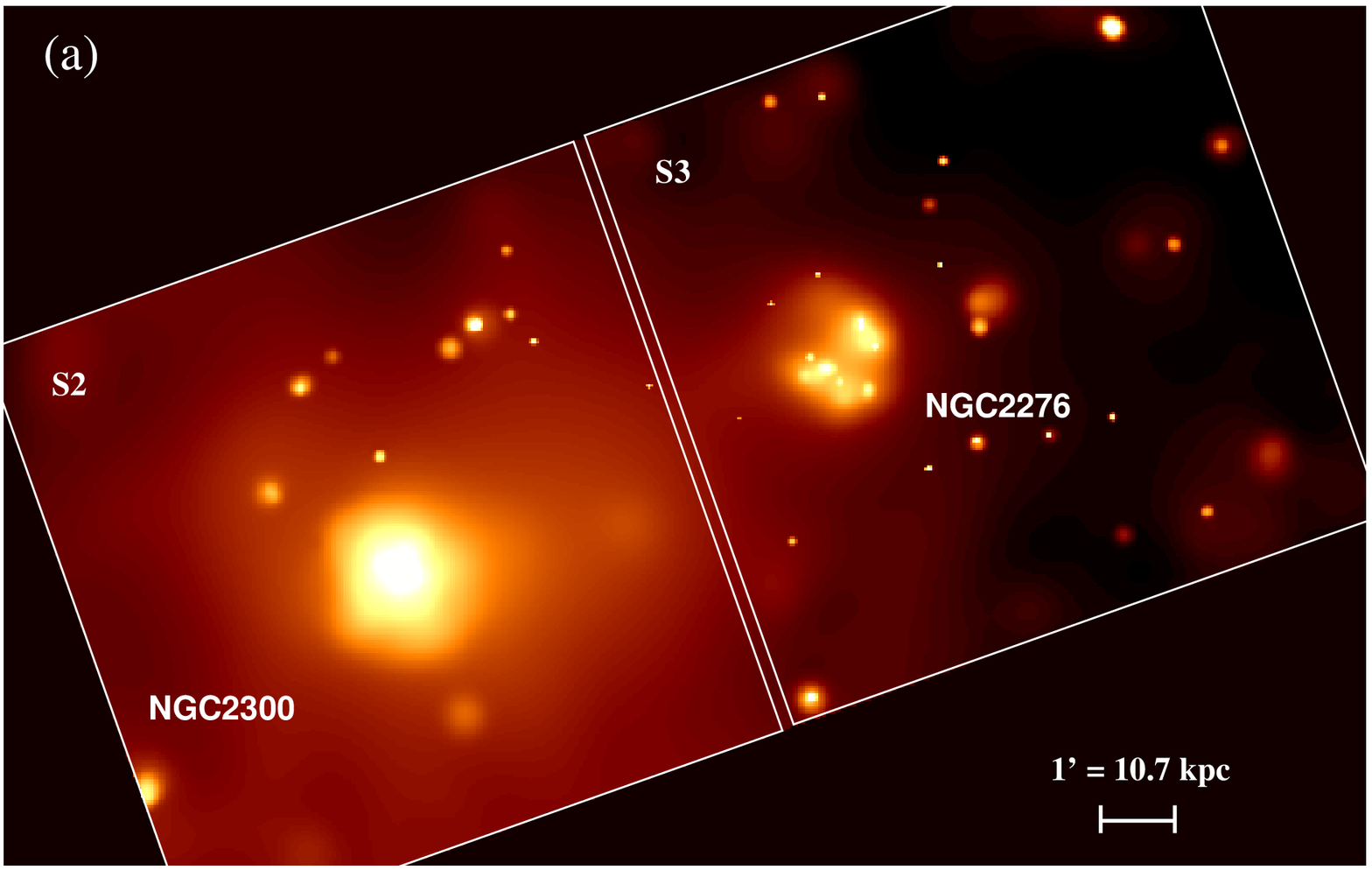}
 \mbox{\hspace{-0.1cm}
   \includegraphics[width=70mm]{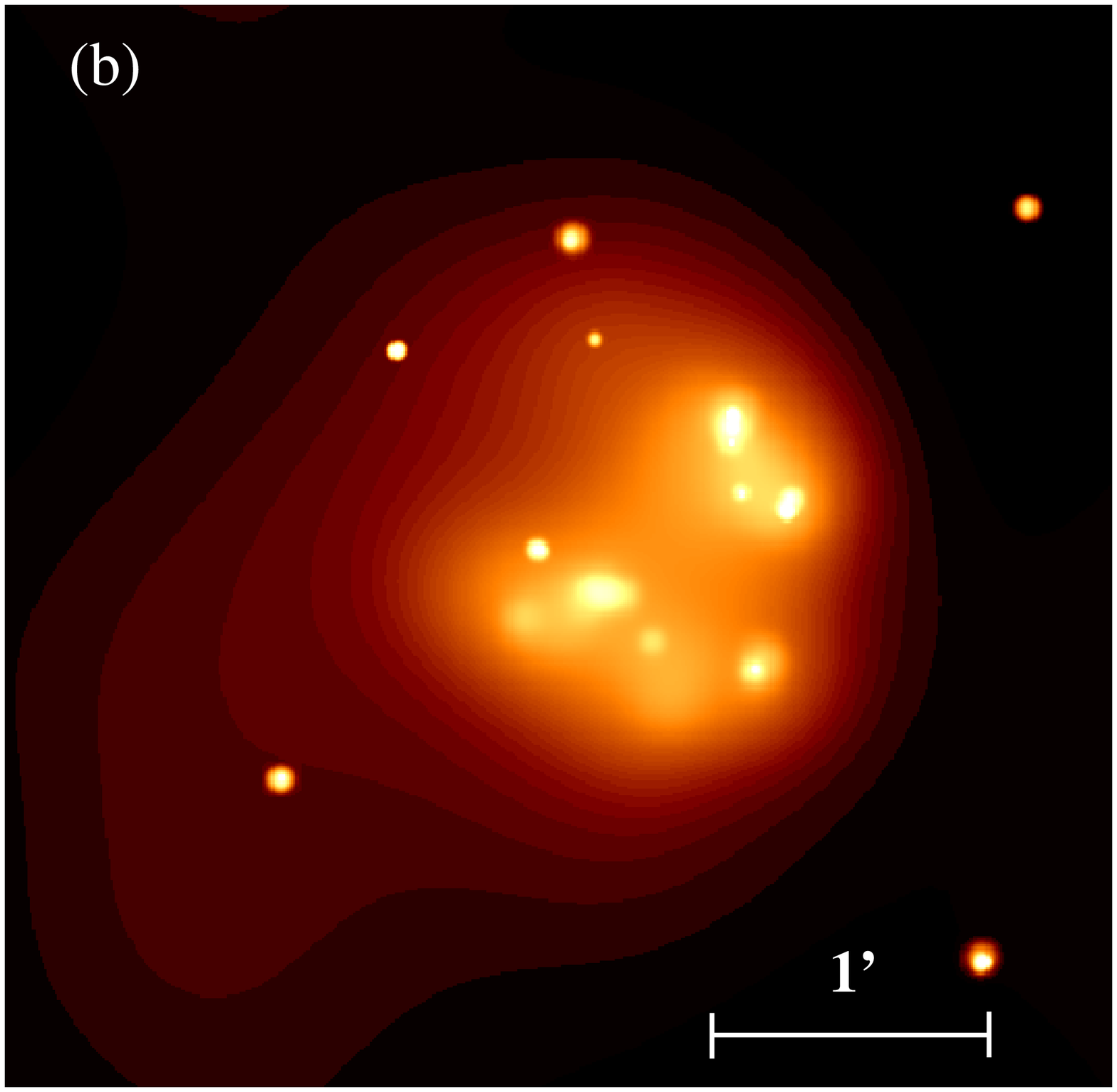}\hspace{-1mm}
\includegraphics[width=68.5mm]{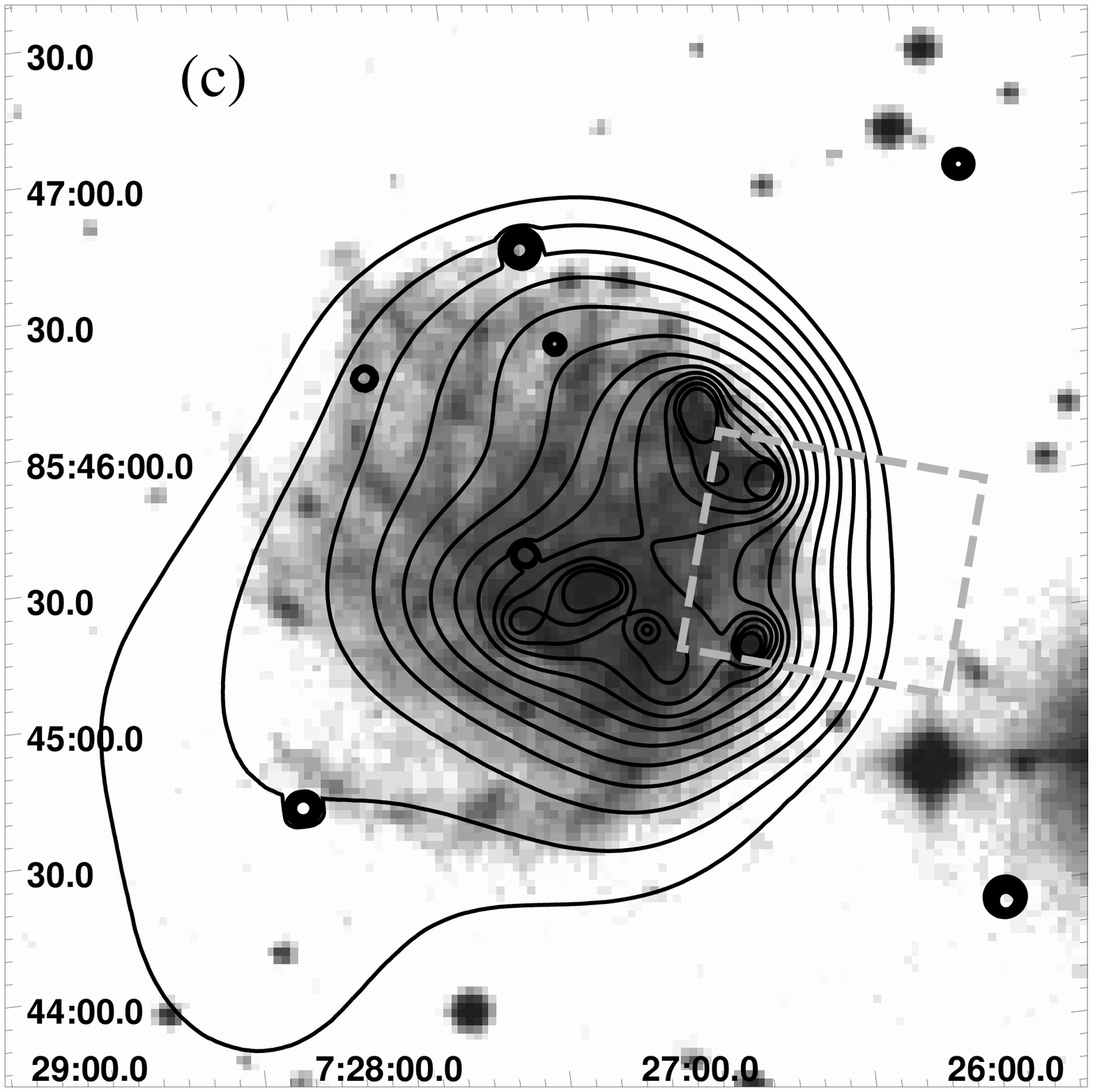}}
\caption{(a) 0.3--2 keV adaptively smoothed image of the full S2 and
  S3 CCDs ($3\sigma$--$5\sigma$ smoothing significance). (b) The
  central $4\times 4$~arcmin$^2$ region surrounding NGC~2276. (c)
  Corresponding X-ray contours overlayed on a Digitized Sky Survey
  image. Dashed box shows the region used for extraction of surface
  brightness profiles across the western edge.}
 \label{fig,N2276}
\end{center}
\end{figure*}

All spectra were accumulated in bins containing at least 25 net counts
and fitted in {\sc xspec} v11.3 assuming the solar abundance table of
\citet{grev98}. For the absorbing component we assumed the Galactic
value of $N_{\rm H}=5.5\times 10^{20}$~cm$^{-2}$ from \citet{dick90}.
Spectral response files were generated using the {\sc mkacisrmf} tool
in {\sc ciao} v3.2, and all response products were weighted
iteratively by a model of the incident source spectrum.

We employed two methods for generating background spectra.  As IGM
emission is expected to cover the S3 chip and thus must be subtracted
from source spectra, background spectra for NGC~2276 were estimated
from source-free regions on the same chip, described below.  Since
vignetting differences between the adopted source and background
regions are small ($\sim$~3~per~cent at the peak spectral energy of
$\sim 0.7$~keV for the region covered by the NGC~2276 disc), results
are in any case likely to be dominated by statistical uncertainties.
However, in cases where the properties of the IGM emission itself
are of interest, we used blank-sky background data to generate
background spectra. These data were scaled to match the 10--12~keV
count rates of our source data in off-source regions on the S3 chip.
Since the assumed value of $N_{\rm H}$ of our source data is a factor
of $\sim 4$ larger than the exposure-weighted mean value of the
relevant blank-sky files, the background is likely to be overestimated
at low energies with this method. Consequently, when using blank-sky
data for background estimation, we restricted spectral fitting to the
0.7--5~keV band, within which no systematic residuals were seen.

The regions used for spectral extraction are shown in
Fig.~\ref{fig,spec}, and a summary of fit results is provided in
Table~\ref{tab,spec}. A detailed investigation of the properties of
hot gas in the elliptical NGC~2300 will be presented elsewhere.

\begin{figure}
\begin{center}
  \includegraphics[width=80mm]{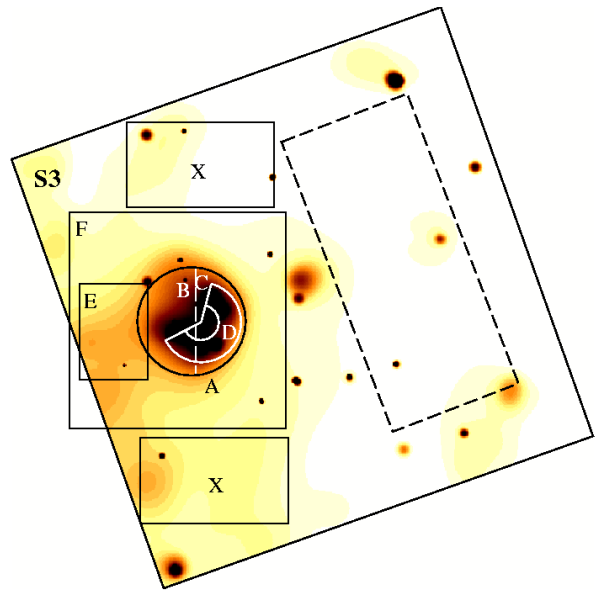}
  \caption{Regions used for spectral fitting of NGC~2276. For clarity,
    excluded point source regions are not shown. Dashed box outlines
    the region used for extraction of on-chip background spectra.}
  \label{fig,spec}
\end{center}
\end{figure}

\section{Results}

\subsection{NGC~2276}

In the 0.3--5 keV band, roughly 1500 net counts are detected from
diffuse emission in the NGC~2276 disc (inside region~A in
Fig.~\ref{fig,spec}).  The X-ray contours of the diffuse emission, as
revealed by Fig.~\ref{fig,N2276}c, are clearly compressed at the
western edge, demonstrating, as already indicated by earlier {\em
  ROSAT} HRI data \citep{davi97}, that also the {\em hot} gas in the
disc displays a disturbed morphology reminiscent of a bow-shock. A 
low--surface-brightness tail extending to the south-east is also visible.
Smoothing the data to lower significances did not reveal evidence of
any additional features in the diffuse emission.

The smoothed image also shows evidence for diffuse emission extending
beyond the western optical edge of the galaxy. To investigate this in
more detail, a surface brightness profile for the unsmoothed diffuse
emission was extracted across the edge, inside the region shown as a
dashed box in Fig.~\ref{fig,N2276}c. The result is plotted in
Fig.~\ref{fig,surfbright}, revealing a steep decline in X-ray surface
brightness which coincides with the edge of the optical disc.
Interestingly, there is indeed evidence for excess emission
immediately outside this edge, extending to a few kpc from the optical
disc. We will return to this feature in Section~\ref{sec,vel}.

\begin{figure}
\hspace{-8mm}
 \includegraphics[width=95mm]{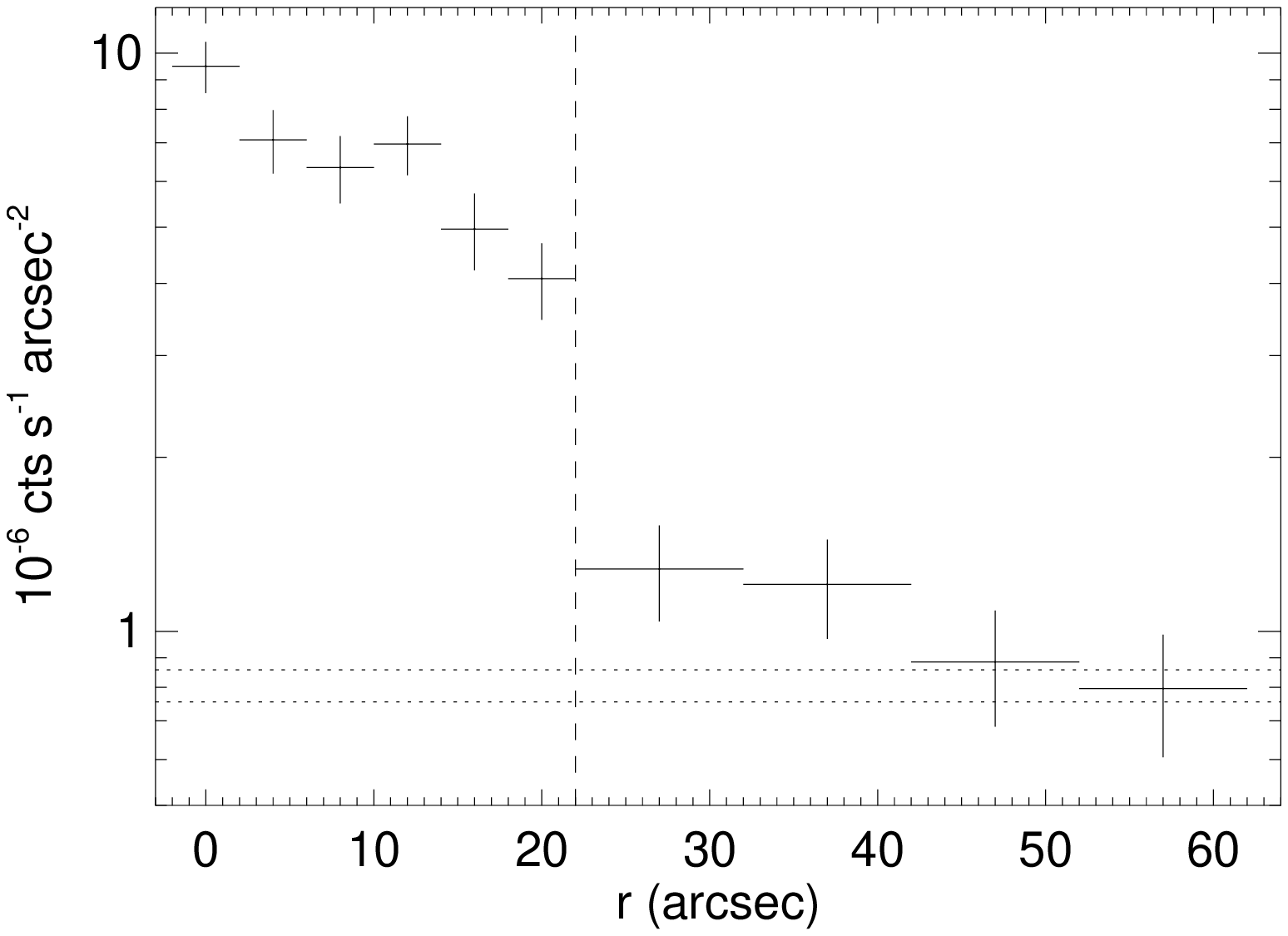}
 \caption{Exposure-corrected 0.3--2 keV surface brightness profile of
   diffuse emission across the western edge of the NGC~2276 disc.
   Dotted lines mark the 90 per cent confidence limits on the
   background level as determined from an on-chip background spectrum
   (i.e.  including emission from the ambient IGM). Vertical dashed
   line marks the extent of the optical disc, quantified as $D_{25}$,
   the ellipse outlining a $B$-band isophotal level of
   25~mag~arcsec$^{-2}$.}
 \label{fig,surfbright}
\end{figure}

For the spectral analysis, the background was evaluated inside the
dashed box shown in Fig.~\ref{fig,spec}.  Since this region is further
from NGC~2300 and hence the IGM emission peak \citep{davi96} than
NGC~2276 itself, any gradient in the IGM surface brightness would
imply that we could be underestimating the background at the position
of NGC~2276. To test this, background spectra were also extracted from
smaller regions in the immediate vicinity of NGC~2276 (regions~X in
Fig.~\ref{fig,spec}), but the spectral results for the disc emission
were found to be indistinguishable at the $1\sigma$ level for all
fitted parameters.

A single-temperature fit to the disc emission (region A in
Fig.~\ref{fig,spec}) was found to be only marginally acceptable.
Given the vigorous star formation in the disc, there could be a
significant contribution from a population of unresolved high-mass
X-ray binaries, or, as seen in many other starburst galaxies, evidence
for a second thermal plasma component. We therefore also tried adding
a power-law or a second thermal component, which both produced better
fits ($\Delta \chi^2 =13.5$ and 17.5, respectively, for 50 degrees of
freedom).  We take the two-temperature model to be a representative
parametrization of the integrated spectrum of diffuse emission. The
total 0.3--2~keV luminosity of this emission is $L_{\rm
  X}=1.86\pm0.37\times 10^{40}$~erg~s$^{-1}$, and its spectrum and
best-fitting model are shown in Fig.~\ref{fig,n2276spec}a.  We tested
whether the temperature derived for the hotter component was affected
by residual emission from either low-- or high-mass X-ray binaries,
fixing this residual component to either a $\Gamma=1.5$ power-law or a
7~keV thermal bremsstrahlung model \citep*{irwi03} for low-mass X-ray
binaries, or a $\Gamma = 1$ power-law for high-mass ones (e.g.
\citealt{vand05}).  The best-fitting temperature for the hotter gas
component is in all cases consistent with the value listed in
Table~\ref{tab,spec}, and the fitted models suggest that unresolved
X-ray binary emission accounts for only 10--15~per~cent of the total
0.3--2~keV flux. For the stellar mass of NGC~2276 listed in
Table~\ref{tab,gals}, this emission level is consistent with the value
of $L_{\rm X}\approx 2\times 10^{39}$~erg~s$^{-1}$ expected for a
population of low-mass X-ray binaries \citep{gilf04}.
\begin{figure*}
\begin{center}
\mbox{\hspace{-2mm}
  \includegraphics[width=70mm]{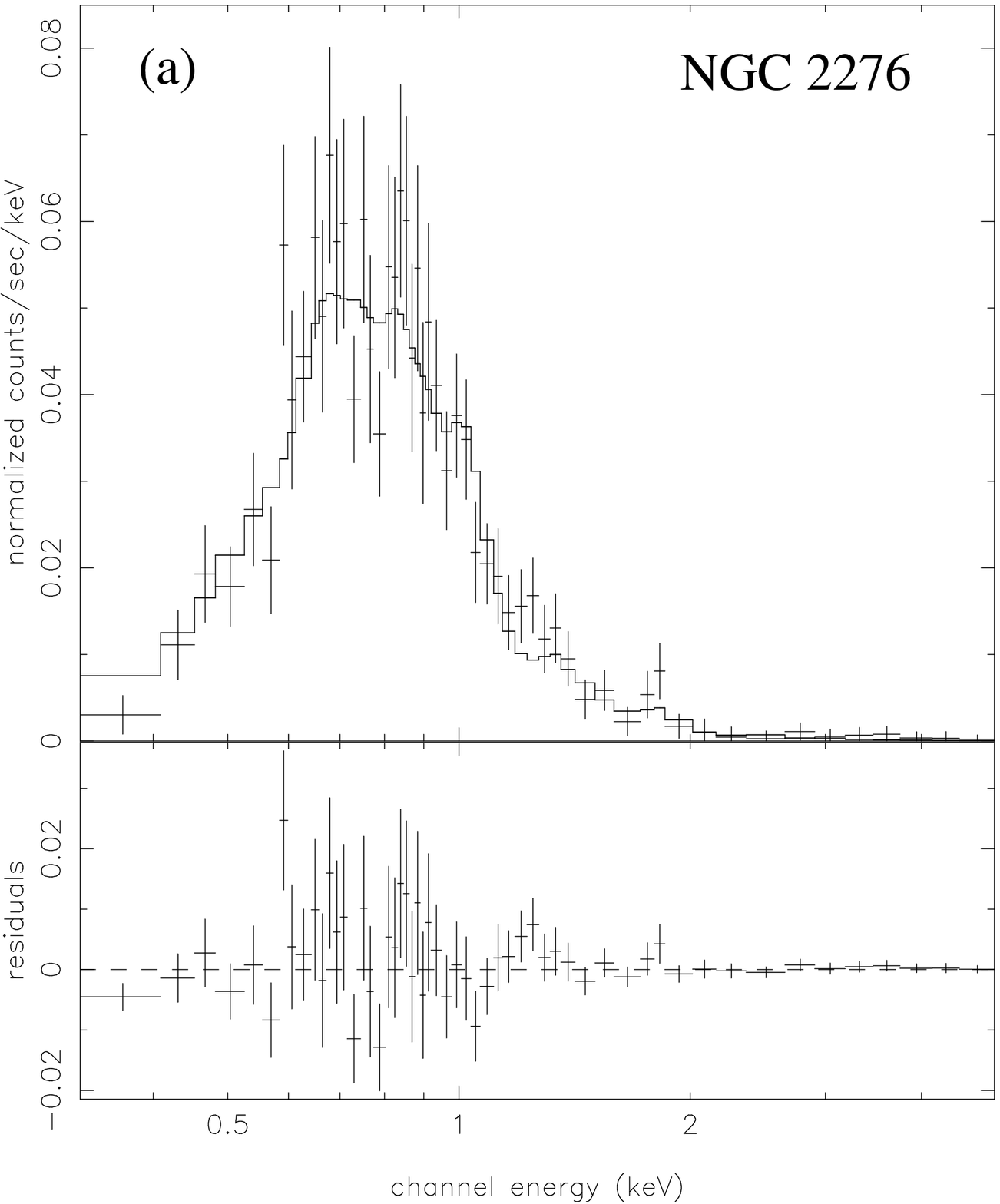}\hspace{22mm}
 \includegraphics[width=70mm]{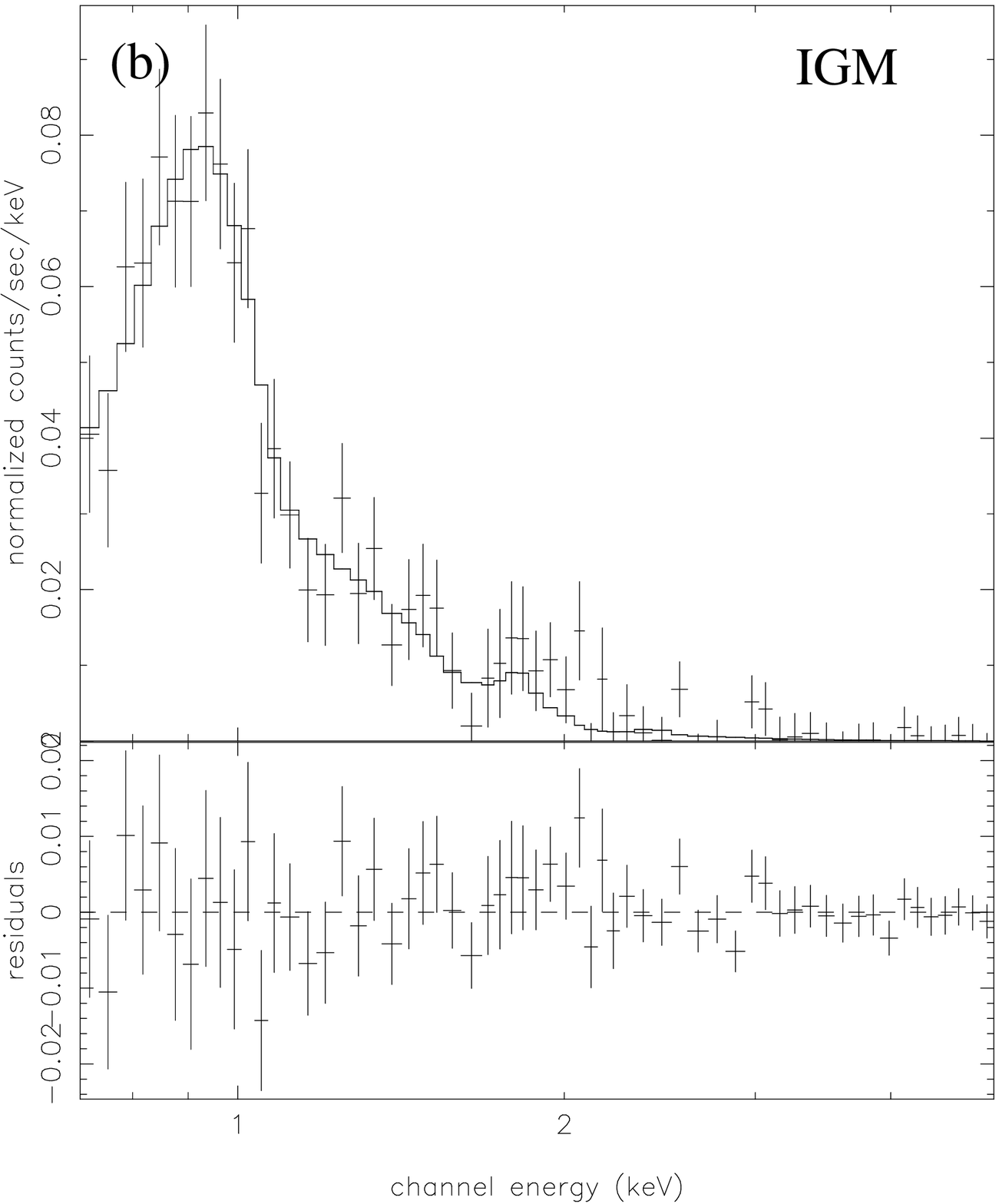}}
\caption{Spectrum and fitted model for (a) the NGC~2276 disc emission 
  and (b) the surrounding IGM on the S3 chip.  Bottom panels show
  fit residuals.}
\label{fig,n2276spec}
\end{center}
\end{figure*}

Spectra were also extracted for the east and west side of the disc
separately (regions~B and C in Fig.~\ref{fig,spec}, selected so as to
contain a similar number of net counts, $\sim 750$ in 0.3--5 keV).
Fitting results are virtually identical for the two regions and are
consistent with those derived for the X-ray bright W edge of the
galaxy alone (region~D). There is thus no spectral evidence that the
hot gas in the W side of the disc is preferentially affected by
external pressure and/or mixing with IGM gas arising from the galactic
motion through the IGM.

We note that the low abundances derived for the disc gas could be
misleading, due to the well-known Fe-bias, arising when fitting the
spectrum of a multi-phase gas with a single-temperature model
\citep{buot00}.  For example, when allowing $Z$ for the
two-temperature model to vary, the result is consistent with
supersolar abundances. This is testimony to the (anticipated)
complexity of the spectrum, and the fitted models should be considered
convenient means of parametrizing the disc spectrum rather than a
complete description of the physical state of hot gas in the disc.

\subsection{IGM}\label{sec,IGM}

In order to validate the properties of the IGM emission as derived
from earlier {\em ROSAT} studies, we modelled the 0.3--2~keV surface
brightness distribution on the full S3~chip alone, using an elliptical
$\beta$--model on top of a constant background, with NGC~2276 excised
out to a radius of 1.5~arcmin. The appropriate exposure map was
included in the fit, and the assumed peak of the emission was taken to
be the position reported by \citet{davi96}. We find
$\beta=0.41^{+0.06}_{-0.09}$, $r_c=4.0\pm 1.8$~arcmin, and $S_0 =
3.1\pm 0.5 \times 10^{-3}$~counts~s$^{-1}$~arcmin$^{-2}$, in excellent
agreement with the {\em ROSAT} result of \citet{davi96} (when
correcting $S_0$ for the difference in effective area).  With a
best-fitting eccentricity of $\epsilon = 0.28^{+0.12}_{-0.15}$ and a
major axis position angle consistent with an east-west orientation,
the derived large-scale morphology of the IGM emission is also
qualitatively similar to the one found by \citet{mulc93} and
\citet{davi96}.

For the spectral analysis, source-free regions on the S2+S3 chips were
defined, and spectra and responses were extracted separately for each
chip. The spectra were fitted both individually for the two chips and
jointly, with the spectrum from the S3 CCD and its best-fitting model
shown in Fig.~\ref{fig,n2276spec}b. A joint fit yields
$T=0.85\pm0.04$~keV and $Z=0.17^{+0.07}_{-0.05}$~Z$_\odot$. The
derived temperature is in good agreement with the value of
$T=0.97^{+0.11}_{-0.08}$~keV obtained by \citet{davi96} from three
combined {\em ROSAT} PSPC exposures of the group. For comparison to
this earlier study, we derive $Z=0.11^{+0.04}_{-0.03}$~Z$_\odot$ using
the abundance table of \citet{ande89}.  This is also consistent with
the result of \citet{davi96}, hence confirming the relatively low
abundance of this group.  The good agreement with the spatial and
spectral results of \citet{davi96} implies that we can justifiably
adopt their values for the peak IGM density and total IGM mass.

\begin{table*}
\begin{minipage}{176mm}
  \caption{Summary of spectral fits to diffuse emission in the 0.3--5
    keV band (0.7--5 keV for the IGM fits, see text). Region labels
    are shown in Figs.~\ref{fig,spec} and \ref{fig,mach}
    ('S3$\setminus$F' means the S3 chip 'excluding F').  Unabsorbed
    X-ray luminosities for the major components are presented in the
    0.3--2~keV band.  Last column gives the goodness of fit for the
    number of degrees of freedom $\nu$.}
\label{tab,spec}
 \begin{tabular}{@{}lllcr}
   \hline
   Region & {\sc xspec} model & Results  & $L_{\rm X}$ & $\chi^2/\nu$\\
   &                   & ($T$/keV, $Z$/Z$_\odot$) & ($10^{39}$~erg~s$^{-1}$) & \\
   \hline
   NGC~2276 disc (A)& {\sc wabs(mekal)} & 
   $T = 0.53^{+0.08}_{-0.05}$, $Z=0.10^{+0.04}_{-0.03}$  
   & \ldots & 58.9/52 (1.13) \\
   \ldots              & {\sc wabs(mekal+pow)} &
   $T=0.34^{+0.04}_{-0.03}$, $Z=2.11^{+25.29}_{-0.61}$, 
   $\Gamma=2.00^{+0.40}_{-0.20}$ & \ldots & 45.4/50 (0.91) \\
   \ldots              & {\sc wabs(mekal+mekal)} &
   $T_1 = 0.32^{+0.03}_{-0.03}$, $T_2 = 1.30^{+1.47}_{-0.35}$ 
   $Z=0.43^{+0.82}_{-0.16}$ & $18.6\pm 3.7$ & 41.4/50 (0.83) \\
   NGC~2276 E disc (B)         & {\sc wabs(mekal)} &           
   $T = 0.51^{+0.11}_{-0.09}$, $Z=0.06^{+0.04}_{-0.03}$ 
   & $10.0\pm 2.9$ & 26.9/28 (0.96)  \\
   NGC~2276 W disc (C)         & {\sc wabs(mekal)} & 
   $T=0.53^{+0.09}_{-0.08}$, $Z=0.11^{+0.08}_{-0.05}$       
   & $9.1\pm 3.0$ & 18.1/25 (0.72) \\
   NGC~2276 W edge (D)         & {\sc wabs(mekal)} &
   $T = 0.55^{+0.07}_{-0.07}$, $Z=0.11^{+0.08}_{-0.04}$ 
   & $13.3\pm 4.0$ & 24.7/24 (1.03) \\
   NGC~2276 ``tail'' (E)   & {\sc wabs(mekal)} & $T=0.84^{+0.35}_{-0.22}$, $Z=0.17$ (fixed)                & $1.9\pm 0.5$ & 0.3/3 (0.11)  \\
   \ldots                         & {\sc wabs(mekal)} & $T=0.76^{+0.20}_{-0.18}$, $Z=1.0$ (fixed)          & \ldots & 0.9/3 (0.30)  \\
   IGM, S2 chip              & {\sc wabs(mekal)} & 
   $T=0.95^{+0.07}_{-0.09}$, $Z=0.27^{+0.19}_{-0.13}$
   & $41.7 \pm 14.6$ & 29.9/46 (0.65) \\
   IGM, S3 chip (S3$\setminus$F) & {\sc wabs(mekal)} & 
   $T=0.81^{+0.05}_{-0.06}$, $Z=0.17^{+0.13}_{-0.07}$
   & $29.5 \pm 9.8$ & 41.5/76 (0.55) \\ 
   IGM, S2+S3             & {\sc wabs(mekal)} & 
   $T=0.85^{+0.04}_{-0.04}$, $Z=0.17^{+0.07}_{-0.05}$ &
   $70.3\pm 14.1$ & 128.1/123 (1.04) \\ 
   Shocked IGM (G)    & {\sc wabs(mekal)} & $T=1.08^{+0.70}_{-0.33}$, $Z=0.17$ (fixed) &
   \ldots & 0.4/3 (0.14) \\ 
   Mach cone (H)      & {\sc wabs(mekal)} & $T=1.06^{+0.32}_{-0.19}$, $Z=0.17$ (fixed) &
   \ldots & 8.1/13 (0.73) \\ 
   Mach cone (G+H)    & {\sc wabs(mekal)}  & $T=1.20\pm 0.27$, $Z=0.17$ (fixed) & 
   \ldots & 8.9/15 (0.59)  \\  
   Unshocked IGM (I)  & {\sc wabs(mekal)}  & $T=0.78\pm 0.10$, $Z=0.17$ (fixed) & 
   \ldots & 38.6/37 (1.04) \\
   \hline
\end{tabular}
\end{minipage}
\end{table*}

\section{Discussion}

The compression of the X-ray isophotes along the west edge of NGC~2276
and the presence of a tail of gas extending towards the east suggest
that the hot gas morphology of the galaxy could be affected by its
motion through the IGM.  However, given the position of NGC~2300 and
hence the IGM peak, it seems worthwhile first to establish to what
extent the appearance of the eastern tail is distorted by
superposition on the broader IGM surface brightness gradient. To
investigate this, we smoothed the derived 2-D IGM model using the same
spatial scales as used for producing Fig.~\ref{fig,N2276}, and
subtracted it from the latter. The result, presented in
Fig.~\ref{fig,mach}, shows that the X-ray tail persists, still showing
an overall morphology in line with the expectation if the gas
distribution of the galaxy is affected by ram pressure caused by
motion towards the west. Also shown in Fig.~\ref{fig,mach} are
contours outlining the radio continuum emission in the galaxy
\citep{davi97}, indicating the presence of a radio tail coincident
with that seen in X-rays. We will now proceed to investigate whether
the overall X-ray and radio morphologies could indeed be indicative of
the direction of motion of NGC~2276 in the plane of the sky.

\begin{figure}
\begin{center}
\includegraphics[width=80mm]{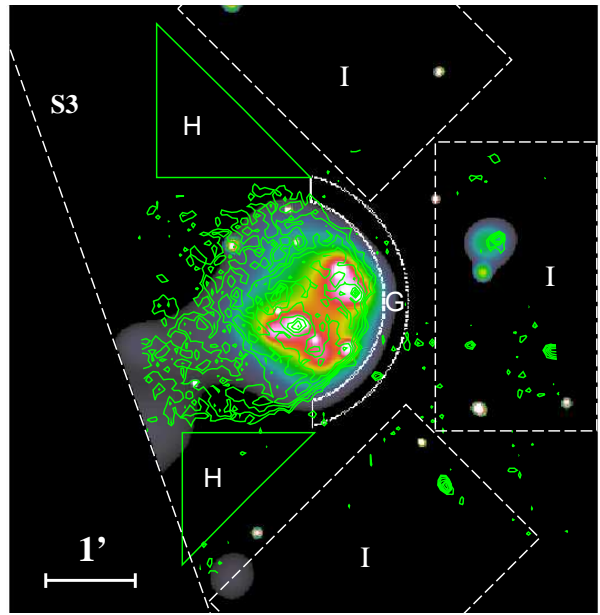}
\caption{As Fig.~\ref{fig,N2276}, but for a $7\times 7$~arcmin region
  around NGC~2276 and with IGM emission subtracted.  Also shown are
  the regions (G+H) outside the W edge and inside the Mach cone used
  to search for evidence of shock-heating of the IGM (see
  Section~\ref{sec,vel} for details), with dashed regions (I) used for
  comparison spectra. Overlayed are Very Large Array 1.49~GHz contours
  from the data presented by \citet{davi97}, starting at
  0.05~mJy~beam$^{-1}$ and spaced by a factor $\sqrt{2}$.}
 \label{fig,mach}
\end{center}
\end{figure}

\subsection{Velocity of NGC~2276 through the IGM}\label{sec,vel}

The distance from the IGM centre to the W edge of NGC~2276 is $\sim
70$~kpc.  The gas mass and $\beta$--profile parameters derived by
\citet{davi96} imply an IGM density outside the W edge of $n_e = 6\pm
0.6 \times 10^{-4}$~cm$^{-3}$. Using $T_{\rm IGM}\approx 0.81$~keV
from Table~\ref{tab,spec}, this then implies a thermal IGM pressure of
$P_{\rm IGM} \approx 1.6\times 10^{-12}$~dyn~cm$^{-2}$ outside this
edge, and a local sound speed $v_s \approx 500$~km~s$^{-1}$.

The radial velocity difference $\Delta v_r$ between NGC~2276 and
NGC~2300, the two most massive group galaxies, is $\Delta v_r \approx
420$~km~s$^{-1}$ (Table~\ref{tab,gals}). The more massive elliptical
NGC~2300 is close to the IGM centre, so $\Delta v_r$ should be a
reasonable measure of the radial velocity of NGC~2276 relative to the
ambient IGM, showing that the true 3-D speed of this galaxy could
easily be supersonic. In the absence of magnetic fields and assuming
that radiative cooling of post-shock gas can be neglected (this will
be justified below), the Rankine-Hugoniot shock adiabat for a
one-dimensional shock would then imply
\begin{equation}\label{eq,RH1}
\frac{n_2}{n_1} = \frac{\gamma+1}{\gamma -1+2/M_1^2},
\end{equation}
\begin{equation}\label{eq,RH2}
  \frac{T_2}{T_1} = \frac{(1-\gamma+2\gamma
    M_1^2)(\gamma-1+2/M_1^2)}{(1+\gamma)^2},
\end{equation}
for the density and temperature ratios of upstream (subscript 1) and
downstream (subscript 2) gas\footnote{As is customary in the
  literature, we use subscripts 1 and 2 to distinguish upstream from
  downstream quantities throughout this paper.}.  Here $M_1$ is the
Mach number of the galaxy's motion through the IGM, and $\gamma$ is
the adiabatic index.  Note that equation~(\ref{eq,RH1}) implies a
maximum density jump of $n_2/n_1=4$ for a shocked $\gamma=5/3$ gas.
Also note that both equations are strictly only valid for the region
immediately in front of the galaxy, where the gas velocity can be
taken to be perpendicular to the shock surface.  For regions along the
shock front further away from the shock symmetry axis, the gas will
encounter an oblique shock wave, inclined by an angle $\phi < 90\degr$
with respect to the galaxy velocity vector. In this case $M_1$ should
be replaced by $M_1{\rm sin}\,\phi$ in equations~(\ref{eq,RH1}) and
(\ref{eq,RH2}) (see, e.g., chapter~9 in \citealt{land87}).

Is the diffuse X-ray emission from the W edge itself arising from a
(shock)-compressed gas component?  A low metal abundance is found for
gas in the W disc, consistent with the value derived for the ambient
IGM. This could point towards gas in the W disc being compressed IGM
material rather than, e.g., starburst generated gas, but the Fe bias
could be important here, as the full-disc spectral fits indicate the
presence of at least two gas phases.  If assuming a face-on disc
(\citealt{grue93} constrain the inclination to
$9^{\circ}<i<17^{\circ}$) of thickness 2~kpc, the derived spectrum and
flux of $5.9\times 10^{-14}$~erg~cm$^{-2}$~s$^{-1}$ imply a mean
electron density of $n_e \approx 0.018$~cm$^{-3}$, a thermal pressure
of $P_{\rm th}\sim 2 n_e T \approx 3.1\times 10^{-11}$~dyn~cm$^{-2}$,
and a gas mass of $M\approx 1.5\times 10^8$~M$_\odot$ for the western
half of the disc.  The corresponding values for the east disc are very
similar, implying a total hot gas mass of $M\approx 3.7\times
10^8$~M$_\odot$ and no large-scale hot gas flows inside the disc due
to pressure imbalances.  Adopting the single-temperature model of the
disc emission, the lower $T$ relative to the IGM value suggests that
this emission is not dominated by a {\em shocked} IGM. The density
jump between IGM and disc gas of a factor $\sim 30$ is also much
larger than can be achieved in a non-radiative shock, thus reinforcing
this conclusion.  As the temperature of the disc gas is typical of
that inferred for starburst galaxies and since, for that temperature,
both the X-ray luminosity and mass of this gas fall right on to the
trends derived for starburst galaxies \citep{read01}, emission inside
the W edge is most likely dominated by that of starburst-generated
gas, although some mixing with the ambient IGM may have taken place.

Fig.~\ref{fig,N2276}c suggests evidence of X-ray emission beyond the
optical W edge of NGC~2276. This component appears spatially distinct
from the hot interstellar medium (cf.\ Fig.~\ref{fig,surfbright}), and
could arise predominantly from a compressed or even shocked IGM.  A
spectral fit to the emission immediately outside the W optical edge
(region~G in Fig.~\ref{fig,mach}, corresponding to the two bins with
excess emission in Fig.~\ref{fig,surfbright}), gives $T =
1.08_{-0.33}^{+0.70}$~keV (for $Z$ fixed at the global IGM value of
0.17~Z$_\odot$ derived for the S3 chip; best-fitting value is $\sim
0.5$~Z$_\odot$ but statistics are too poor to obtain useful
constraints on this quantity and neither the best-fitting $T$ nor its
errors are very sensitive to changes in the assumed $Z$).  Although
nominally higher, this temperature is still consistent with that of
the ambient IGM, but is inconsistent with the single-component value
for the interstellar medium (ISM) in the disc.  Moreover, comparing
blank-sky and on-chip background levels suggests that only one-third
of this emission is from fore-/background IGM. Motivated by this, we
plot in Fig.~\ref{fig,pressure} the thermal pressure corresponding to
the surface brightness plot shown in Fig.~\ref{fig,surfbright},
assuming that most of this emission immediately outside the W edge
arises within a line-of-sight depth of $\la 20$~kpc ($\approx 2r_{\rm
  D}$, where $r_{\rm D}$ is the disc radius).  Intriguingly, there is
evidence for a build-up of pressure immediately outside the optical
disc, consistent with a scenario in which the IGM in front of the
galaxy is being pressurized due to the motion of the galaxy. For the
assumed 20~kpc depth, we find a local mean density of $n_e \approx
1.6\pm 0.7\times 10^{-3}$~cm$^{-3}$ and a pressure $P_{\rm th}\approx
4.8\times 10^{-12}$~dyn~cm$^{-2}$ in this region.  This suggests a
density jump of a factor $n_2/n_1 = 2.6\pm 1.2$ with respect to the
ambient IGM, indicating a shock, but a corresponding jump in $T$
cannot be unambiguously established from these results alone.

\begin{figure}
\begin{center}
  \mbox{\hspace{-6mm}
    \includegraphics[width=93mm]{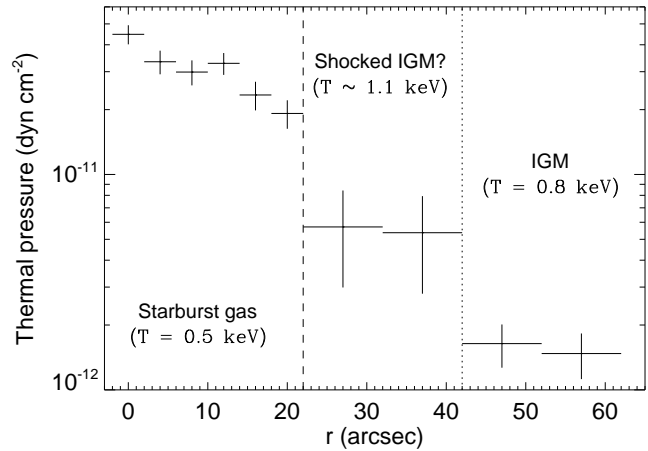}}
\caption{NGC~2276 pressure profile across the W edge.}
\label{fig,pressure}
\end{center}
\end{figure}

If a shock front is present, gas immediately inside the Mach cone will
be affected also at some distance from NGC~2276. The asymptotic Mach
angle $\alpha$ (the cone opening angle with respect to the symmetry
axis) should satisfy sin\,$\alpha = M_1^{-1}$. An initial guess of
$\alpha \approx$~40--45${\mbox{\degr}}$, based on the X-ray and
optical morphology of the W edge, would correspond to a mild shock of
Mach number $M_1\approx 1.5$.  To help constrain the temperature of
gas inside this potential shock front, we also considered regions
outside the optical disc but immediately inside the $M_1\approx 1.5$
Mach cone (regions~H in Fig.~\ref{fig,mach}). Despite the fact that
these regions are further away from the shock symmetry axis than the W
edge and so should experience a smaller temperature jump (as mentioned
above, due to $M_1$ in equation~\ref{eq,RH2} being replaced by
$M_1{\rm sin}\,\phi$, where $\alpha<\phi<\pi/2$), we find
$T=1.06^{+0.32}_{-0.19}$~keV, which is still marginally inconsistent
with the unshocked IGM value.  In an attempt to maximize the S/N ratio
and obtain tighter constraints on the post-shock temperature, we
performed a joint fit to the emission in these regions and that of the
W edge (i.e.\ regions G+H in Fig.~\ref{fig,mach}). We readily
acknowledge the fact that $T$ is not expected to be uniformly raised
across this region, but the result is expected to be dominated by
statistical errors anyway.  This approach gives $T = 1.20 \pm
0.27$~keV, thus strengthening the inconsistency with the local IGM
value.  The temperature jump between the putative shocked and
unshocked regions is then $T_2/T_1 = 1.48\pm 0.28$, in agreement with
the expectation from equation~(\ref{eq,RH2}) for an $M_1\approx
1.5$~shock. For the derived density jump of $n_2/n_1 = 2.6\pm 1.2$,
this result is also in formal agreement with the maximum ratio
$T_2/T_1 =2.6_{-1.3}^{+3.8}$ predicted by equations~(\ref{eq,RH1}) and
(\ref{eq,RH2}) for $\phi = 90\degr$.  As a consistency check, we
confirmed that the emission level in various regions upstream of the
$M_1\approx 1.5$ Mach cone (regions~I) is consistent with that of the
(background plus) local IGM, rendering it unlikely that our results
for the two regions inside the Mach cone are substantially affected by
inaccurate background subtraction.

One would also expect an increase in surface brightness across the
edges of the Mach cone, but such a jump is not convincingly detected.
However, this could be due to geometrical effects. In particular, we
suffer from not knowing the exact 3-D direction of motion of NGC~2276
and the resulting orientation of the Mach cone, which makes it
non-trivial to provide a meaningful surface brightness plot for the
Mach cone corresponding to that presented in Fig.~\ref{fig,surfbright}.
Also, when projected on to the sky, the surface brightness will be
continuous at a shock even though the gas density is discontinuous at
the front, because shock fronts are invariably curved. Projection
effects will therefore tend to smear out any sharp increase in the
X-ray emission across the front.  Nevertheless, even after subtracting
the best-fitting IGM surface brightness model from the data, we do
detect a clear enhancement in surface brightness inside the Mach cone
(regions~H in Fig.~\ref{fig,mach}), which is found to be $\sim
80$~per~cent higher than in regions immediately outside this cone
(regions~I). There is a corresponding jump in $T$, measured to be
$T=0.78\pm 0.10$~keV in regions~I, compared to the $1.20 \pm 0.27$~keV
inside the Mach cone.

The data are thus consistent with a mildly shocked IGM.  The density
jump observed across the W edge, and the temperature jump in this and
the Mach cone regions, translate, via equations~(\ref{eq,RH1}) and
(\ref{eq,RH2}), into consistent Mach numbers $M_1 = 2.4_{-1.1}^{+1.8}$
and $M_1 = 1.5\pm 0.3$, respectively.  Combining the information on
density and temperature, one can evaluate the shock (and hence galaxy)
speed directly as (see e.g. \citealt*{henr04})
\begin{equation}\label{eq,vgal}
  v_{\rm gal} = \frac{1}{1-n_1/n_2}\left[\frac{kT_1}{\mu m_p}
    \left(\frac{n_2}{n_1}-1 \right)\left(\frac{T_2}{T_1}-\frac{n_1}{n_2}\right)
  \right]^{1/2} ,
\end{equation}
yielding $v_{\rm gal} = 784\pm 103$~km~s$^{-1}$. The derived error on
$v_{\rm gal}$ is the dispersion resulting from 10,000 Monte Carlo
realizations of equation~(\ref{eq,vgal}), for which parameters were
drawn from a Gaussian distribution centred at their best-fitting
values and having $\sigma$ equal to the measured $1\sigma$
uncertainty.  Nominally, the result implies a Mach number $M_1 =
1.54\pm 0.20$.

However, several effects tend to make the measured value of the
temperature jump smaller than it would be in a simple, normal shock at
the velocity of the galaxy.  First, emission from unshocked gas
projected into the line-of-sight could reduce the measured temperature
of the shocked gas. We have attempted to reduce the impact of this
effect by using on-chip background estimates from unshocked regions
(see Fig.~\ref{fig,spec}).  Second, in region~G in
Fig.~\ref{fig,mach}, the shock strength should vary quite
significantly, since, as mentioned above, the effective Mach number
varies as sin\,$\phi$, where $\phi$ is the angle between the galaxy
velocity vector and the shock front.  Thirdly, for the same reasons,
the shock is almost certainly weaker for region~H than in region~G.
The fact that we can even detect a temperature excess in region~H
almost certainly requires that the galaxy is moving considerably
faster than our simple estimate.  Fourthly, the postshock pressure
returns quickly to the ambient gas pressure in the downstrean flow.
Most of the gas begins to expand adiabatically and cool immediately
behind the shock (gas close to the stagnation point at the leading
edge of the galaxy is the exception).  In combination, these effects
will generally reduce the measured temperature jump. The derived jump
will therefore almost certainly underestimate the true Mach number of
the galaxy relative to the IGM.

In addition, the actual Mach number is further underestimated due to
projection effects. Given a line-of-sight velocity $\Delta v_r$ of
NGC~2276 relative to the group, a sound speed $v_s$, and the apparent
(i.e.\ measured) opening angle $\alpha$ of the Mach cone projected on
to the sky, the intrinsic Mach number $\cal{M}$ is given by
\begin{equation}
  {\cal M} =  \frac{ [ 1+ (\Delta v_r \mbox{sin\,}\alpha)^2/v_s^2]^{1/2}}
  {\mbox{sin\,} \alpha}.
\end{equation}
Using $\Delta v_r\approx 420$~km~s$^{-1}$ and $\alpha \approx
$~40--45${\mbox{\degr}}$, this corresponds to $ {\cal M} \approx 1.7$.
If adopting the fractional errors derived for equation~(\ref{eq,vgal})
and adding these in quadrature to those resulting from varying
$\alpha$ within the range 40--45\degr, one finds $ {\cal M} = 1.70\pm
0.23$, i.e.\ a pre-shock velocity of $v_1 = 865\pm 120$~km~s$^{-1}$ of
the galaxy relative to the IGM.  We note, for the various reasons
mentioned above, that the actual speed of the galaxy could be even
higher, although this is not easily quantified and may well be
subsumed by the large uncertainties.  For the purposes of this paper,
which is to investigate the importance of ram pressure, the
'deprojected' Mach number of $ {\cal M} \approx 1.7$ should be a
conservative choice which we will consequently adhere to in the
following. The implied density and temperature jumps from
equations~(\ref{eq,RH1}) and (\ref{eq,RH2}) are then $n_2/n_1=2.0$ and
$T_2/T_1=1.7$, respectively.

\subsection{The effects of ram pressure}\label{sec,ram}

Assuming a shock is present, conservation of the mass flux density
$\rho_{\rm ICM} v_{\rm gal}$ perpendicular to the shock front implies
that the true ram pressure felt by the galaxy will be smaller than
that upstream of the front. For ${\cal M}= 1.7$, the implied density
increase of a factor 2.0 corresponds to a decrease in galaxy--IGM
velocity to $v_{\rm gal} \approx 430$~km~s$^{-1}$. This is thus the
velocity $v_2$ of post-shock gas immediately behind the shock front.
We note that the cooling time of this gas is $\sim 14$~Gyr, thus
justifying the non-radiative assumption underlying
equations~(\ref{eq,RH1}) and (\ref{eq,RH2}).  The resulting ram
pressure $P_{\rm ram}=\rho_{\rm ICM} v_{\rm gal}^2 = \rho_2 v_2^2
\approx 3.7\times 10^{-12}$~dyn~cm$^{-2}$, which, combined with the
increased thermal IGM pressure behind the shock front, yields a total
IGM pressure $P_{\rm tot} = P_{\rm ram} + P_{\rm th} \sim 9\times
10^{-12}$~dyn~cm$^{-2}$.  This is a factor of 3 less than the mean
thermal pressure of the hot ISM in the disc, at first suggesting that
the morphology of hot disc gas would be only marginally affected by
ram pressure.

However, a more detailed treatment of the effects of ram pressure, in
particular the question of ram-pressure {\em stripping}, must take
into account two additional considerations. One is the gravitational
restoring force per unit area of the disc \citep{gunn72}, $F/A = 2\pi
G \Sigma_\ast \Sigma_{\rm g}$, where $\Sigma_\ast$ and $\Sigma_{\rm
  g}$ are the surface densities of stars and gas, respectively. The
classical condition for ram-pressure stripping to occur then becomes
$\rho_{\rm ICM} v_{\rm gal}^2 > 2\pi G \Sigma_{\ast} \Sigma_{\rm g}$,
i.e.\
\begin{equation}
  \left(\frac{n_{e}}{\mbox{cm$^{-3}$}}\right)\left(\frac{v_{\rm gal}}
    {\mbox{km/s}}\right)^2
  \ga 100
  \left(\frac{M_\ast}{\mbox{M$_{\odot}$}}\right) \left(\frac{M_{\rm g}}
    {0.1 M_\ast} \right) \left(\frac{r_{\rm D}}{\mbox{kpc}}\right)^{-4},
  \label{eq,gg}
\end{equation}
where $n_{e}$ is the IGM electron density, $r_{\rm D}$ the disc
radius, and $M_\ast$ and $M_{\rm g}$ is the mass of stars and gas in
the disc, respectively.  The validity of this estimate has been
supported by results of numerical simulations \citep{abad99}. The
other issue to consider is that while the ram pressure itself is
reduced at the shock, $P_{\rm th}+\rho v^2$ remains conserved (for an
inviscid fluid).  Thus, the reduction in ram pressure behind the shock
front is balanced by in increase in static (thermal) pressure.  Now,
the difference between the static pressure at the stagnation point,
where the flow meets the leading edge of the galaxy, and the upstream
thermal pressure is approximately equal to the upstream ram pressure.
As long as the galactic disc remains intact (and the effective
Reynolds number is reasonably high), a region of separated flow can be
expected behind the disc, where the pressure is approximately equal to
the downstream (i.e.\ ambient) gas pressure.  The quantity relevant to
the issue of ram pressure stripping, the force per unit area acting on
a disc normal to the flow, is thus close to the pre-shock ram
pressure.

Using therefore the pre-shock values of $n_e=n_1=6\times
10^{-4}$~cm$^{-3}$ and $v_{\rm gal}=v_1 = 865$~km~s$^{-1}$ in
equation~(\ref{eq,gg}), along with $M_\ast$ from Table~\ref{tab,gals},
$r_{\rm D}=10$~kpc, and $M_{\rm g}=6.4\times 10^{9}$~M$_\odot$
\citep{youn89}, one finds a ram-pressure falling short of $F/A$ by a
factor $\sim 15$. The central IGM density is only a factor 1.7 larger
than the value assumed, so using this value instead does not change
the immediate conclusion that ram pressure stripping should not have
been an important factor in the recent evolution of NGC~2276.
Moreover, the assumed value of $n_e$ could be optimistic, as NGC~2276
could be further away from the IGM centre than indicated by the
projected distance.  However, the velocity dispersion of the group is
$\sigma \approx 300$~km~s$^{-1}$ \citep{mulc93}, in excellent
agreement with the expectation from the $\sigma$--$T$ relation for
X-ray bright groups \citep{osmo04}.  The larger values of both $\Delta
v_r$ and $v_{\rm gal}$ may suggest that NGC~2276 is currently passing
through the group core, so we are probably not making a large error in
taking the projected distance to NGC~2300 as representative of the
true distance to the IGM density peak.

An important caveat associated with the stripping criterion of
equation~(\ref{eq,gg}) is that the gas is assumed to be held in place
by a homogeneous disc thin enough to be described solely by its
surface density $\Sigma$. In order to provide a more realistic picture
of the effects of ram pressure, the calculation should be repeated as
a function of radius $R$ in the disc midplane, and vertical disc
height $|z|$, under suitable assumptions about the distribution of
gas, stars, and dark matter (DM).  For this purpose, we set up a model
of the gravitational potential $\Phi(R,z)$ of NGC~2276.  We adopted
the model of \citet{abad99} which consists of a spherical bulge with
density profile
\begin{equation}
  \rho_{\rm b}(r)=\frac{M_{\rm b}}{2\pi r_{\rm b}^2} 
  \frac{1}{r(1+r/r_{\rm b})^3},
\end{equation}
a spherical DM halo,
\begin{equation}
  \rho_{\rm h}(r)=\frac{M_{\rm h}}{2\pi^{3/2}} \frac{\eta}{r_{\rm t} 
    r_{\rm h}^2} \frac{\mbox{exp}(-r^2/r_{\rm t}^2)}{(1+r^2/r_{\rm h}^2)},
\end{equation}
and exponential stellar and gaseous discs, each of the form
\begin{equation}
  \rho_{\rm d}(R,z) = \frac{M_{\rm d}}{4\pi R_{\rm d}^2 z_{\rm d}} 
  \mbox{exp}(-R/R_{\rm d})\mbox{sech}^2(z/z_{\rm d}).
\end{equation} 
Here $M_{\rm b}$, $M_{\rm h}$, and $M_{\rm d}$ are the total masses of
each component, $r_{\rm b}$ and $r_{\rm h}$ are the scalelengths of
the bulge and halo, respectively, $r_{\rm t}$ is the halo truncation
radius, $R_{\rm d}$ is the cylindrical scalelength of the disc
components and $z_{\rm d}$ the corresponding thickness, and finally
\begin{equation}
\eta = \{{1 - \pi^{1/2}q\mbox{\,exp}(q^2)[1-\mbox{erf}(q)]}\}^{-1},
\end{equation}
where $q=r_{\rm h}/r_{\rm t}$ and erf is the error function.  We adopt
again the stellar mass $M_\ast$ from Table~\ref{tab,gals}, a cold gas
mass $M_{\rm g}=6.4\times 10^9$~M$_\odot$, a hot X-ray emitting gas
mass $M_{\rm hot} = 3.7\times 10^8$~M$_\odot$ (cf.\
Section~\ref{sec,vel}), and a bulge-to-disc stellar mass ratio of
0.25, appropriate for an Sc spiral like NGC~2276.  In order to comply
with the observed size of the disc, the scalelength for the disc
components have been fixed by the requirement that 90~per~cent of all
gaseous and stellar mass should lie within $r_{\rm D}\approx 10$~kpc.
A vertical scaleheight of 1~kpc was adopted for the X-ray emitting
component, representative of the values derived for galaxies of the
mass and size of NGC~2276 in the disc galaxy sample of \citet{stri04}.
The mass and truncation radius of the DM halo are then set by the
requirement that the maximum disc circular velocity $v_{\rm rot}$ be
within the observed limits of $v_{\rm rot} \approx
120^{+55}_{-40}$~km~s$^{-1}$ (mean value from the HyperLeda database).
The adopted model parameters are listed in Table~\ref{tab,model}, and
Fig.~\ref{fig,vrot} illustrates the contribution of the various model
components to the rotational disc velocity; the adopted model has a
maximum rotational velocity of $v_{\rm rot}\approx 140$~km~s$^{-1}$ at
$r\approx 7$~kpc, consistent with observations.

\begin{table}
\begin{center}
  \caption{Adopted parameters of the galaxy model. $L$ is the
    characteristic scalelength for each component, i.e.\ $r_{\rm h}$
    and $r_{\rm t}$ for the halo, $r_{\rm b}$ for the bulge, and
    $R_{\rm d}$ and $z_{\rm d}$ for the disc components.}
\label{tab,model}
 \begin{tabular}{@{}lcl}
   \hline
   Component     & $M_{\rm total}$ & $L$        \\ 
                 & ($10^{10}$~M$_\odot$) & (kpc) \\ \hline
   DM halo       &  8.00           & 3.5, 30 \\
   Stellar bulge &  0.55           & 0.3 \\
   Stellar disc  &  2.15           & 3.0, 0.25 \\         
   Cold gas disc &  0.64           & 2.5, 0.25 \\
   Hot gas disc  &  0.04           & 3.0, 1.0 \\
   \hline
\end{tabular}
\end{center}
\end{table}

\begin{figure}
\begin{center}
\mbox{\hspace{-7mm}
\includegraphics[width=93mm]{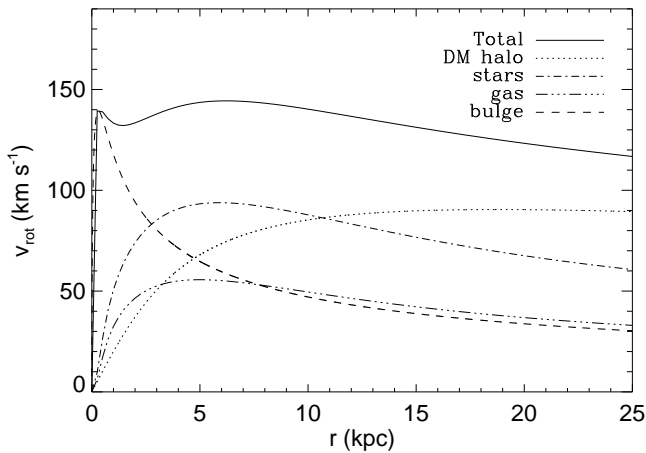}}
\caption{Contribution to the rotational disc velocity from the
  different components of the galaxy model.}
\label{fig,vrot}
\end{center}
\end{figure}

The restoring gravitational acceleration $\frac{\partial
  \Phi}{\partial z}(R,z)$ in the direction $z$ perpendicular to the
disc can be evaluated analytically for each model component using the
equations of \citet{abad99}, to whom we refer for more details. Also,
the values of $\Delta v_{\rm r}$ and (the pre-shock) $v_{\rm gal}=v_1$
for NGC~2276 imply motion at an angle $\xi \approx 29^\circ$ with
respect to the plane of the sky. Given the small inclination under
which the galaxy is viewed, this implies a reduction in the nominal
ram pressure felt by the disc by a factor $\approx 4$ relative to a
situation where the disc is experiencing a face-on encounter with the
IGM. The condition for ram-pressure stripping now becomes
\begin{equation}
 \Sigma_{\rm g}\left(\frac{\partial \Phi_{\rm b}}{\partial z} + 
 \frac{\partial \Phi_{\rm h}}{\partial z} +
 \frac{\partial \Phi_{\rm g}}{\partial z} +
 \frac{\partial \Phi_{\ast}}{\partial z} \right)
 < \rho_1 (v_1\mbox{sin}\xi)^2 .
\label{eq,strip}
\end{equation}
In Fig.~\ref{fig,strip}, we plot the left-hand side of
equation~(\ref{eq,strip}) along with the derived ram pressure. As was
also found by \citet{roed05} for their galaxy models, the
gravitational restoring force in the outer disc ($R\ga 5$~kpc) is seen
to be nearly independent of $z$ for all interesting values of this
parameter.  The figure suggests that at present, ram pressure alone is
probably not powerful enough to remove large amounts of cold disc gas.
But it also shows that the inferred stripping radius, outside which
$P_{\rm ram} > F/A$, is close to 10~kpc for the adopted galaxy model,
in excellent agreement with the observed truncation radius of cold and
hot gas along the W edge of NGC~2276.  Ram pressure could therefore
have played an important role in establishing the present size of the
gas disc.

Fig.~\ref{fig,strip} further illustrates the region outside which
$P_{\rm ram}$ exceeds the thermal pressure of the cold ($T\la 10^4$~K)
ISM not immediately affected by star formation.  Outside this region,
one would expect the distribution of cold gas, such as the H$\alpha$
emitting disc gas, to be significantly modified by ram pressure.
Regarding the hot X-ray emitting ISM, $P_{\rm ram}$ exceeds the
thermal pressure of this gas at heights $|z|\ga$~1--2~kpc above the
disc.  Starburst outflows would therefore be strongly affected by ram
pressure and could be swept away before having time to fall back on to
the disc.  Furthermore, if the X-ray emitting disc gas is mainly
generated in stellar outflows, its distribution would at least
initially follow that of young stars and hence that of the H{\sc i}
out of which these stars are formed (recall that the H{\sc i}
distribution is also compressed along the W edge).  Such outflows will
expand along the steepest density gradient, which will be
perpendicular to the disc and thus close to a direction along the line
of sight. When viewed in projection, the X-ray emission from stellar
outflows may therefore retain the apparent compression along the W
edge both at early and late outflow stages.

\begin{figure}
\begin{center}
\mbox{\hspace{-4mm}
\includegraphics[width=92mm]{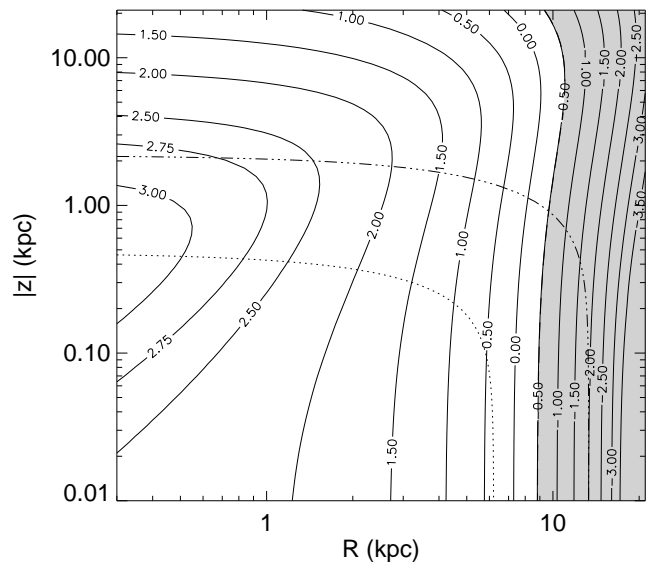}}
\caption{Logarithmic isocontours of the gravitational restoring force
  $F/A$ perpendicular to the disc, in units of
  $10^{10}$~M$_\odot$~kpc$^{-3}$~(km s$^{-1}$)$^2$.  The shaded area
  outlines the region where galactic gas can be stripped by the
  derived ram pressure, log\,$P_{\rm ram} \simeq -0.5$.  Dotted and
  dash-dotted lines mark the regions outside which $P_{\rm ram}$
  exceeds the thermal pressure of cold and hot disc gas,
  respectively.}
\label{fig,strip}
\end{center}
\end{figure}

It is instructive to also consider the effects of ram pressure on the
hot coronal gas believed to surround at least massive ($v_{\rm rot}\ga
120$~km~s$^{-1}$) spiral galaxies at low redshift
\citep{whit91,toft02,birn03}. This gas is assumed to provide the
reservoir from which large disc galaxies are continuously replenishing
the disc gas consumed in star formation, and is essential to
explaining the extended star formation histories of isolated disc
galaxies like the Milky Way. For a galaxy similar to NGC~2276, this
gas component is expected to have $T \la 0.1$~keV and $L_{\rm X}\la
10^{38}$~erg~s$^{-1}$ \citep{toft02,rasm06} and so would not be
detectable in the present data. With an expected warm/hot gas mass of
$\sim 10^8$~M$_\odot$ outside the disc \citep{rasm06}, the mean
thermal pressure of this gas will be $P\la 3\times
10^{-13}$~dyn~cm$^{-2}$ for a spherical corona of radius $r_{\rm
  D}\approx 10$~kpc (log $P\la -1.35$ in the units of
Fig.~\ref{fig,strip}).  While Fig.~\ref{fig,strip} indicates that a
considerable fraction of this gas would remain bound in the galactic
gravitational potential, ram pressure should disrupt its gradual
inflow in this potential, preventing it from attaining the densities
necessary for rapid cool-out on to the disc (see \citealt{acre03}).
Effectively, this will eventually halt star formation in the disc in a
scenario identical to strangulation, but there is currently sufficient
gas in the NGC~2276 disc for star formation to proceed at its present
rate for at least another Gyr.

Transport processes could be more efficient than ram-pressure in
removing gas from the galaxy. The expected mass loss rate due to
viscous stripping via Kelvin-Helmholtz instabilities is $\dot M_{\rm
  KH} \approx \pi r_{\rm D}^2 \rho_{\rm ICM} v_{\rm gal}$
\citep{nuls82}, i.e.\
\begin{equation}\label{eq,KH}
  \dot M_{\rm KH} \approx 0.1 \left(\frac{n_e}{\mbox{cm$^{-3}$}}\right)
  \left(\frac{r_{\rm D}}{\mbox{kpc}}\right)^2  
  \left(\frac{v_{\rm gal}}{\mbox{km s$^{-1}$}}\right) 
  \mbox{ M$_\odot$ yr$^{-1}$} . 
\end{equation}
This would imply $\dot M_{\rm KH} \approx 5$~M$_\odot$~yr$^{-1}$ for
the disc of NGC~2276.  Note that, because $\rho_{\rm ICM} v_{\rm gal}$
is conserved across the shock front, the expected efficiency of
viscous stripping is unaffected by the presence of a shock front.  The
viscous stripping efficiency is also largely insensitive to the
orientation of the galaxy with respect to the direction of motion
\citep{nuls82,quil00}, unlike that of ram pressure stripping.

The mass loss rate due to thermal conduction of heat from the IGM to
the ISM, whereby the colder disc gas is heated above the escape
temperature of the galaxy (only counteracted by radiative cooling),
can be 3.5 times larger than $\dot M_{\rm KH}$ \citep{nuls82}. But the
efficiency of this process is uncertain due to the strong magnetic
fields present in the disc, as evidenced by the radio continuum
emission (indicating a field strength roughly twice the average of
spirals; \citealt{humm95}). The fact that the large majority of the
ISM ($\sim 95$~per~cent) is in a cold phase suggests that conduction
of heat from the IGM is not currently an important process.

In summary, these results indicate that shock compression can explain
the presence of slightly denser and hotter gas outside the leading
edge of galaxy, and that it could certainly exert a significant
influence on the ISM, both in terms of establishing the size of the
gas disc and compressing the H{\sc i} and X-ray contours along the W
edge, as well as helping to trigger the substantial star formation
along this edge. In this context it is also interesting to note the
optical result of \citet{davi97} that the radial gradient in [O{\sc
  iii}]/H$\beta$ is twice as steep on the western side compared to the
eastern side of the disc, a result which could be attributed to gas in
the W disc being swept inward (potentially promoting the growth of a
larger bulge). Starburst outflows would be swept back by ram pressure,
and the cooling out of hot coronal gas on to the disc would be
inhibited. Ram-pressure, however, is probably insufficient to strip
large amounts of gas from the disc, whereas hydrodynamic instabilities
arising along the ISM/IGM boundary could be removing ISM material from
the galaxy at a rate of several solar masses per year.

\subsection{The nature and origin of the X-ray tail}

The presence of an X-ray tail extending eastwards from the disc of
NGC~2276 is established at the $3\sigma$ significance level, based on
the enhancement in surface brightness within this region relative to
the on-chip background. While its presence is thus tentative rather
than conclusive, the location of the tail matches that of a similar
structure seen in both H{\sc i} and radio continuum data
\citep{humm95, davi97}.  Only one clear-cut example of such an X-ray
tail trailing a spiral in a poor group has so far been reported
\citep{mach05a}. The X-ray luminosity of the NGC~2276 tail is $\sim
2\times 10^{39}$~erg~s$^{-1}$, about 10~per~cent of that of the
diffuse disc gas. The tail is not seen in the optical; in particular,
there is no evidence from H$\alpha$ data of ongoing star formation in
this region \citep{grue93}. A single X-ray point source is detected at
$>3\sigma$~significance in the tail, but statistics are too poor
($\sim 20$ net counts) to allow a test of whether this is a background
source.

The hot tail gas has a temperature $T \sim 0.8$~keV, consistent with
that of the surrounding IGM. Extending to a galactocentric distance of
$\sim 20$~kpc, the tail gas should be essentially unbound from the
galaxy, for which the characteristic escape temperature at this
position is $kT_{\rm esc}= \mu m_p v_{\rm esc}^2/3 \approx 0.1$~keV as
derived from our galaxy model.  If the tail line-of-sight depth is
$\sim r_{\rm D} \approx 10$~kpc, and assuming $Z=0.2$~Z$_\odot$, its
density and hot gas mass are $n_e \sim 2.8\times 10^{-3}$~cm$^{-3}$
and $\sim 1.2\times 10^8$~M$_\odot$, respectively, with both these
numbers reduced by roughly a factor 2 for a $Z=$~Z$_\odot$ plasma.
The galaxy moves a distance corresponding to the extent of the tail
($\sim 10$~kpc at $v_2 = 430$~km~s$^{-1}$) in $\sim 20$~Myr.  If the
entire mass of the X-ray tail represents unbound gas displaced from
the disc due to the motion of NGC~2276, this would imply a mass loss
rate averaged over the disc of order $\dot M_{\rm tail} \sim
3$--6~M$_\odot$~yr$^{-1}$. Of course, this relies on the assumption
that the speed of the tail (relative to the galaxy) equals that of the
shocked gas.  However, as we will see, a significant fraction of tail
gas is likely to represent hot disc gas which has been heated and
accelerated by the shock, supporting this assumption.

For the H{\sc i} tail, a rough estimate of its mass can be obtained
from the H{\sc i} map of \citet{davi97}.  Using $M_{\mbox{H{\sc i}}}
\approx 2.36\times 10^5 (D/{\rm Mpc})^2 (F_{\mbox{H{\sc i}}}/\rm{Jy})$
(where $F_{\mbox{H{\sc i}}}$ is the integrated H{\sc i} flux in the
tail; see, e.g., \citealt{rose02}), we find $M_{\mbox{H{\sc i}}} \sim
4\times 10^7$~M$_\odot$.  Taken at face value, this would correspond
to a mass loss rate of $\sim 2$~M$_\odot$~yr$^{-1}$ of cold gas,
assuming, as for the X-ray tail, that the galaxy has moved a distance
corresponding to the extent of the tail in $\sim 20$~Myr. The H{\sc i}
tail could therefore be accounted for by viscous stripping.  However,
our mass model of the galaxy suggests that this cold gas would remain
bound in the potential of NGC~2276.  As our main focus is the gas that
can potentially be completely lost from the galaxy, we will
concentrate on the origin of the X-ray tail in the following.

There appear to be at least four potential origins for this hot tail:
gravitationally focused IGM material (a ``wake''), gas stripped off by
ram pressure or other transport processes, starburst-driven outflows
being swept back by the same processes, or gas being tidally displaced
by interaction with NGC~2300 and/or the group potential. A robust
metallicity constraint could help distinguish between these scenarios,
but the tail metallicity is essentially unconstrained. One immediately
attractive explanation is that the tail results from viscous
stripping, as the expected mass loss rate due to this mechanism is
consistent with that inferred from the properties of the tail. It is
nevertheless instructive to consider alternative explanations, so
below we will discuss these different scenarios in more detail.

With spectral properties similar to those of the ambient IGM, the tail
could be a wake of gravitationally focused IGM material.  Although in
analytical theory large-scale wakes are unlikely to be produced by
galaxies moving supersonically \citep{sake00}, the simulation results
of \citet{stev99} indicate that wakes {\em can} be created behind
supersonic (group) galaxies. If so, they should be most easily
observable in a relatively cool ICM \citep{stev99,sake00}.  However,
in order to produce an observable wake, the accretion radius for
Bondi-Hoyle accretion must exceed the galaxy radius \citep{sake00}
which translates into the requirement
\begin{equation} 
  M_{\rm gal} \geq 8.7\times 10^{11}\left(\frac{T}{\mbox{keV}}\right)
  \left(\frac{r_{\rm D}}{10 \mbox{ kpc}}\right) \mbox{ M$_\odot$}
\end{equation}
for the total galaxy mass, i.e.\ $M_{\rm gal}\ga 8\times
10^{11}$~M$_\odot$ for the $r_{\rm D}\approx 10$~kpc disc of NGC~2276.
Given a stellar mass of $M_\ast \la 3\times 10^{10}$~M$_\odot$
(Table~\ref{tab,gals}), it seems unlikely this requirement can be met
unless NGC~2276 is enveloped in an unusually massive dark matter halo.
This can probably be ruled out by our rotation curve modelling, as the
upper limit to $v_{\rm rot}$ of 175~km~s$^{-1}$ suggests $M_{\rm gal}
<2.5\times 10^{11}$~M$_\odot$ for our adopted galaxy model. Another
concern is that the tail gas is at least a factor of 2 denser than the
surrounding IGM. For adiabatic accretion behind the galaxy,
conservation of entropy, $S=T/\rho^{2/3}$, would then predict a
corresponding temperature ratio of $T_{\rm tail}/T_{\rm IGM}\approx
1.6$, which is not favoured by the data.  Consequently, while
dedicated hydro-simulations would be necessary to assess the
feasibility of this scenario in detail, we tentatively conclude that
the tail is unlikely to be a wake.

Another explanation for the tail could be sought in a tidal
interaction between NGC~2276 and NGC~2300. From the morphology of
NGC~2276, \citet{humm95} estimate that the time $t_{\rm app}$ since
closest approach between the galaxies is between half and one
revolution of NGC~2276, about $3\times 10^8$~yr. Assuming NGC~2300 is
at rest within the group potential, a lower limit to $t_{\rm app}$ can
be obtained from the tangential velocity $v_t = (v_{\rm gal}^2 -
\Delta v_r^2)^{1/2}$ of NGC~2276, which can be constrained to $v_t \la
870$~km~s$^{-1}$, conservatively using our upper limit of ${\cal M}
\approx 1.9$.  This implies $t_{\rm app} \ga 80$~Myr for a projected
distance of 75~kpc.  Assuming an average galactocentric distance of
$r\approx 15$~kpc, the dynamical time-scale for tail gas to recentre
on NGC~2276 is $t= (2r^3/GM)^{1/2} \approx 160$~Myr as derived from
our galaxy model.  Since this is larger than $t_{\rm app}$, the tail
could potentially be the remnant of an interaction with NGC~2300. An
argument against this, however, is the fact that the mass of hot gas
in the tail constitutes $\sim$ 15--25~per~cent of all X-ray emitting
gas in NGC~2276. If tidal interactions were responsible for the tail,
then a similar fraction of stellar light should be deposited in a
tidal tail.  This is easily ruled out by optical observations. It is
also curious that, if the tail is older than 80~Myr, it has not been
wound up as a result of taking part in the overall differential
rotation of the NGC~2276 disc.  Given $v_{\rm rot}\sim
120$~km~s$^{-1}$, the age constraint implies at least a quarter of a
full revolution since the time of closest approach with NGC~2300.  It
is hard to see how the observed tail morphology could survive for so
long.

A third possibility for the origin of the tail is related to the star
formation rate of NGC~2276 of $\sim$~2--5~M$_\odot$~yr$^{-1}$, as
estimated from the relations of \citet{kenn98} using either the
H$\alpha$ flux of \citet{davi97}, or the far-infrared luminosity
(calculated from {\em IRAS} 60 and 100~\micron~fluxes).  Hence, the
tail could represent gas lost in starburst outflows which have
subsequently been swept back by ram pressure. Given the inclination of
the galaxy, the presence of such outflows cannot be directly inferred
from the present data.  However, in the disc galaxy sample of
\citet{stri04}, all galaxies having a surface supernova (SN) rate
above 40~Myr$^{-1}$~kpc$^{-2}$ also exhibit significant extraplanar
X-ray emission, presumably resulting from starburst outflows. Using
the same approach as \citet{stri04}, we find a corresponding SN rate
of 280~Myr$^{-1}$~kpc$^{-2}$ for NGC~2276. This is comparable to
values found for prototypical starburst outflow galaxies such as
NGC~253 and NGC~1482, strongly suggesting that outflows are also
taking place in NGC~2276.  Furthermore, the kinematics of H$\alpha$
gas in the disc is consistent with gas motion out of the plane of the
galaxy \citep{grue93}, and the observed steepening of the radio
spectral index with frequency is consistent with adiabatic cooling in
a galactic wind \citep{humm95}.

If we assume most of the tail gas originated along the starbursting
western edge where it was instantaneously accelerated to the
post-shock velocity $v_2$, the typical time-scale for this gas to
reach the tail region would be $\sim 45$~Myr. Using the stellar
population synthesis models of \citet{bruz03}, we find that even a
10~M$_\odot$~yr$^{-1}$ starburst of this age can easily be
accommodated by the present optical broad-band colours of NGC~2276.
The resulting mass loss due to SN-driven outflows is not easily
estimated, as it depends not only on the average mass expelled by each
SN, but also on the amount of ISM swept up by the expanding hot
bubbles (the 'mass-loading' of the outflow), and on the fraction of
outflowing gas that can eventually escape the disc.  However,
observational studies of starburst outflows in spiral galaxies suggest
that, at star formation rates below a few times
$10$~M$_\odot$~yr$^{-1}$, the mass outflow rate is comparable to the
star formation rate \citep*{mart99,rupk05}. This would suggest an
upper limit to the outflow mass {\em loss} rate of $\sim
5$~M$_\odot$~yr$^{-1}$ for NGC~2276.  We therefore conclude that a
scenario in which starburst outflows are being swept back by ram
pressure provides a possible explanation for some, potentially all, of
the hot tail gas.

The X-ray tail thus does seem likely to be ISM displaced from the disc
by transport processes.  If the hot tail gas did indeed originate in
the $T\approx 0.5$~keV starburst-generated gas that probably dominates
the disc X-ray emission, one could ask whether any mechanism could
heat hot disc gas from this temperature to the tail temperature of
$T_{\rm tail}\approx T_{\rm IGM}\approx 0.8$~keV within the relevant
time-scales. We can first obtain an order-of-magnitude estimate of the
time-scales involved for conductive heating from the IGM, assuming
heating between the above temperatures over a length-scale of $\sim
5$~kpc (from the projected edge to the centre of the tail).  The
scalelength of the temperature gradient $l_T = T/|\nabla T| \sim
10$~kpc, much longer than the electron mean free path $\lambda_e
\approx 0.3 (T/\mbox{keV})^2 (n_e/10^{-3}\mbox{ cm$^{-3}$})^{-1}$~kpc,
which is at most a few hundred pc.  Heat conduction at the Spitzer
rate \citep{sara88} would then proceed on a time-scale
\begin{equation}
 t_{\rm cond} \sim 0.2 
 \left(\frac{n_e}  { 10^{-3}\mbox{ cm$^{-3}$}}\right) 
 \left( \frac{l_T}{\mbox{kpc}}\right)^2 
 \left( \frac{T}{\mbox{keV}}\right)^{-5/2} \mbox{ Myr}, 
\end{equation}
of order $t_{\rm cond} \sim 100$~Myr. Given the complexity of the
problem, this is only a rough estimate, since gas would be heated
while being displaced from the disc (changing $l_T$), turbulence could
act to mix IGM and disc gas, and magnetic fields in and above the disc
could be suppressing heat conduction by some unknown factor (cf.\ the
presence of cold fronts observed in some clusters of galaxies, see,
e.g., \citealt*{vikh01}). While bearing these caveats in mind, we note
that the galaxy would have moved 5 kpc in only $\sim 10$~Myr,
suggesting that the tail cannot be conductively heated gas as the
time-scale for this is too long.  Further, if the tail were galactic
gas heated by conduction, then magnetic fields should be unimportant
in the tail region, which appears inconsistent with the presence of
significant radio continuum emission (Fig.~\ref{fig,mach}). A more
promising heating mechanism is shock heating, which could raise $T$ by
a factor $\sim 1.7$ (cf.\ equation~\ref{eq,RH2}), consistent with hot
disc gas being heated from $\sim 0.5$ to $\sim 0.8$~keV.  But heating
may, in fact, not be required at all; a small fraction of tail gas
could be the very hot, tenuous component of starburst outflows
predicted by hydrodynamical simulations of such outflows
\citep{stri00}.

A plausible explanation for the X-ray tail is therefore that it
represents hot disc gas originating in supernova outflows, mildly
shocked and removed from the disc by ram pressure.  We note that the
tail is unlikely to be shock-heated {\em cold} disc gas, as the
pressure of the cold ISM is larger than the combined thermal and ram
pressure of the IGM. A collision between a fragment of the cold, dense
ISM and the IGM would form a shock in the latter rather than
significantly heating the cold disc gas.

Summarizing, the X-ray tail indicates a mass loss rate of $\dot M_{\rm
  tail} \sim$~3--6~M$_\odot$~yr$^{-1}$ of hot gas.  The two mechanisms
of viscous stripping and starburst outflows could each be responsible
for the loss of $\sim 5$~M$_\odot$~yr$^{-1}$ of gas, and so can each
explain the amount of hot gas in the tail.  Other processes, such as
tidal interactions or Bondi-Hoyle accretion, do not appear to be
favoured by the data.  The presence of an H{\sc i} tail coincident
with that seen in X-rays suggests that cold gas is also being removed
from the disc at a rate of $\sim 2$~M$_\odot$~yr$^{-1}$, probably by
viscous stripping rather than ram pressure. Our mass model of the
galaxy suggests that this cold gas would remain bound in the potential
of NGC~2276.  This implies that most ($\sim$~60--100~per~cent) of the
gas escaping the galaxy is in a hot phase, the exact fraction
depending on how much of the H{\sc i} in the tail remains bound.  We
note that, owing to the strong dominance of H{\sc i} gas over X-ray
gas in the disc, most of the gas stripped by viscous stripping would
be expected to be in a cold phase.  The overall picture could
therefore be one in which viscous stripping is mainly removing cold
disc gas, whereas ram pressure acting on starburst outflows removes
the hot starburst-generated gas, with some of the stripped gas being
subsequently heated by shocks.  From our X-ray data alone we cannot
constrain the relative importance of these two processes for the hot
gas; the only safe conclusion is that they could have a comparable
impact. It seems possible, however, that star formation could have an
important indirect effect on the efficiency of viscous stripping, by
making the hot disc gas more vulnerable to the effects of
hydrodynamical instabilities. Only dedicated simulations could
quantify the importance of this effect.

\subsection{Importance of tidal interactions}

Evidence of gravitational interactions between NGC~2276 and the nearby
group elliptical NGC~2300 have been reported, suggesting that IGM
interactions may not be solely responsible for the observed features
of this spiral (see \citealt{grue93}; \citealt{davi97}).  The aim of
this section is to address the question of whether tidal interactions
are really necessary to explain the features of NGC~2276, and if so,
whether their importance can be constrained.  This pertains mainly to
the compression of the stellar and gaseous material along the western
edge, as we have just shown that the eastern gas tail is unlikely to
have a tidal origin.

There are essentially three separate claims of evidence for an ongoing
or past tidal interaction between NGC~2276 and NGC~2300: (i)
\citet{forb92} claim that NGC~2300 has a tidal extension, interpreting
this as evidence of a past interaction with NGC~2276; (ii)
\citet{grue93} argue that certain peculiar features of the velocity
field of H$\alpha$ emitting disc gas are hard to explain by ram
pressure and therefore must have a tidal origin; (iii) the fact that
the stellar disc, in particular the older population of stars
dominating the $R$-band light, has a truncated distribution similar to
that of the H{\sc i} and H$\alpha$ emission, suggests that this could
not have been induced by the present ram pressure and so again must
have a tidal origin.  The claims are backed by the more indirect
arguments that ram pressure is probably not strong enough to induce
the observed peculiarities of NGC~2276, such as the truncation of the
stellar and gaseous discs and the enhanced star formation along the W
edge. In line with this, the model results of \citet{fuji99} suggest
that ram-pressure compression of gas cannot lead to an enhancement in
the star formation rate by more than a factor of 2, even for galaxies
falling through cluster centres. If so, the strong star formation of
NGC~2276 may require a tidal origin. Tidal interactions {\em can}
enhance the star formation rate (e.g.\ \citealt{kenn87}), although it
remains unclear whether they can cause an order-of-magnitude increase
as could be the case for NGC~2276 \citep*{miho91}.

However, there are several lines of evidence which suggest that tidal
interactions may not be essential to explaining the properties of
NGC~2276.  Firstly, assuming NGC~2300 is stationary in the group
potential, it is questionable whether tidal interactions are really
very efficient at a relative velocity of $v_{\rm gal} \sim
850$~km~s$^{-1}$.  Secondly, the tidal extension of NGC~2300 reported
by \citet{forb92} is claimed to protrude towards the north-east, and
is thus not in the direction of the present projected position of
NGC~2276.  Thirdly, regarding the peculiar velocity field of H$\alpha$
gas in the disc of NGC~2276, as \citet{grue93} note themselves, the
kinematics of this H$\alpha$ gas can also be explained by gas motion
out of the plane of the galaxy, and it is not strictly necessary to
invoke a tidal origin.  Given the very likely presence of starburst
outflows, this interpretation seems at least as attractive.

Another issue is the question of whether tidal interactions must be
invoked to explain the truncation of the $R$-band light along the W
edge, and whether the strong star formation along this edge can be
associated with such interactions. It is clear that ram pressure
compression of molecular cloud complexes could lead to enhanced star
formation along the W edge.  If a factor $\sim 2$
increase in the star formation rate (SFR) can be obtained this way
\citep{fuji99}, then the SFR prior to the IGM interaction should have
been $\sim $~1--3~M$_\odot$~yr$^{-1}$.  This is not an unreasonably
high value, and observations of galaxies falling into the cluster
A1367 \citep{gava95} suggest that SFR enhancements at even stronger
levels can indeed result from ICM interactions.  Even more pertinent
is the fact that the time since closest approach with NGC~2300 is $\ga
80$~Myr and hence larger than the typical age of massive stars
dominating the $B$-band light. The current high SFR is therefore
unlikely to have been triggered exclusively by tidal interactions.  As
for the truncation of the stellar disc, if young stars are
preferentially formed in regions where external pressure contributes
to the compression of molecular gas, then $B$-band and H$\alpha$ light
should trace the cold gas morphology, as observed. The red and
near-infrared light, arising predominantly from an older stellar
population, could be truncated if the IGM interaction has persisted
for sufficiently long, perhaps several $10^8$~yr.  There is a clear
indication that also the {\em JHK} isophotes are compressed along the
W edge (albeit far less strongly than the blue light).  The position
and derived 3-D velocity of NGC~2276 implies that the galaxy has been
within the group virial radius of $\sim 1$~Mpc \citep{davi96} for at
least $\sim 1$~Gyr and possibly much longer.  It therefore seems
possible that NGC~2276 has experienced considerable ram pressure for
several $10^8$~yr, so a tidal origin for the truncation of $R$-band
light is not necessarily favoured over ram pressure compression.

Tidal interactions should also lead to centrally peaked SFR
enhancements (e.g., \citealt*{keel85,barn96,berg03}), contrary to what
is observed for NGC~2276. Even so, gas should recover from tidal
distortions on a sound crossing time-scale, which, even for a 5~kpc
structure, is only $\sim 10$~Myr at the temperature of the hot ISM.
This is far shorter than the time since closest approach between
NGC~2276 and NGC~2300, so only some long-lasting dynamical excitation
triggered by the interaction could be maintaining the W structure.

From these arguments, it appears that the tidal scenario on its own
faces too many difficulties to be able to explain the morphology of
NGC~2276.  We cannot rule out that gravitational effects have played a
role in shaping the properties of this galaxy, but they seem unlikely
to be dominant.  On the other hand, a combination of ram-pressure
(acting for several tens, perhaps hundreds, of Myr) and starburst
outflows provides a natural explanation for the enhanced star
formation primarily along the W edge, the compression of radio,
H$\alpha$, broad-band optical/near-infrared, and X-ray isophotes along
this edge, the origin and properties of the eastern gas tail, and the
H$\alpha$ kinematics in the disc.

\section{Implications for galaxy evolution in groups}

There is strong evidence (e.g.\ \citealt*{jone00,bick02}) that spirals
in rich clusters of galaxies are transformed into S0 galaxies by
interactions with the cluster environment, a process underlying part
of the well-known morphology--density relation seen in clusters (but
see also \citealt{burs05}).  The fact that such a relation exists also
for groups \citep{hels03} shows that the underlying mechanisms are
also at work in group environments. The relation appears to be
stronger in groups with a detectable hot intragroup medium than in
those without \citep{osmo04}, but this may only reflect an increased
efficiency of the relevant processes in collapsed systems, not
necessarily a connection with the presence of hot, dense gas.

The H{\sc i} deficiency of NGC~2276 mentioned in
Section~\ref{sec,intro} suggests that this galaxy could already have
lost an amount of disc gas comparable to the current H{\sc i} supply
of $6.4\times 10^{9}$~M$_\odot$, and that stripping processes {\em
  are} acting in this system. If the present mass loss rate of $\dot
M_{\rm tail} \approx 3$--6~M$_\odot$~yr$^{-1}$ is representative, then
NGC~2276 could have been losing H{\sc i} for $\sim$~1--2~Gyr
(comparable to the sound crossing time of the group) and will have
exhausted its supply in another $\sim$~1--2~Gyr. This indicates that
galaxies in groups could lose all atomic gas over the course a few Gyr
as a result of gas stripping by the intragroup medium. While this is
much slower than in rich cluster cores, where such processes are
expected from simulation results to take place on time-scales of tens
of Myr \citep{abad99,quil00}, it could still be sufficiently rapid to
have made a significant impact on the galaxy population in most groups
that have now collapsed.  This is particularly true for cluster
galaxies, as these were incorporated into groups first and can have
experienced such interactions for a significant fraction of a Hubble
time. A well-known example, whose origin could be at least partly
explained by such stripping processes, is the H{\sc i} deficient
spirals seen in the outskirts of the Virgo cluster \citep{sanc04}.

Star formation, interaction-induced or not, will contribute to
exhausting the gas supply of spirals; for NGC~2276, the current star
formation rate of $\sim 5$~M$_\odot$~yr$^{-1}$ suggests that
interaction-induced star formation may, at least temporarily, be as
important as stripping itself in exhausting the gas supply of spirals
in dense environments (although some of the gas going into star
formation will be returned to the ISM on time-scales of $\sim
10^7$~yr).  The question remains whether such starburst activity is
needed to facilitate gas loss from group galaxies. This cannot be
unambiguously addressed by the present study, as both the observed
X-ray and H{\sc i} tails of NGC~2276 could be accounted for by viscous
stripping alone.  An indication that starburst outflows are not a {\em
  prerequisite} for mass loss in groups is the fact that certain group
galaxies, such as Holmberg~II \citep{bure02} and NGC~2820
\citep{kant05}, also appear to be losing gas because of ram pressure
or viscous stripping, without showing any signs of starburst activity.
A 22~ks {\em ROSAT} study with a limiting point source sensitivity of
$\sim 10^{37}$~erg~s$^{-1}$ found no diffuse X-ray gas in Holmberg~II
\citep*{kerp02}, implying that any X-ray tail of this galaxy is orders
of magnitude fainter than the $2\times 10^{39}$~erg~s$^{-1}$ found for
the NGC~2276 tail. However, based on its $B$-band luminosity of
$1.3\times 10^9$~L$_\odot$ \citep{tull88} and the relation of
\citet{read01}, Holmberg~II is expected to show a total diffuse X-ray
luminosity of $\sim 2\times 10^{37}$~erg~s$^{-1}$. Any X-ray tail of
this galaxy would therefore likely remain undetected in the {\em
  ROSAT} data.  Furthermore, since we are not aware of any dedicated
X-ray studies of the gas tail in NGC~2820, we cannot reliably compare
the exact extent to which these two galaxies contain hot tail gas with
our corresponding results for NGC~2276.

In the case of a fairly massive spiral like NGC~2276, another
important effect of ram pressure is its ability to prevent any coronal
gas from cooling out on to the disc. This would eventually quench star
formation by cutting off the supply of fresh material for this
process.  Further, since the stripping radius will decrease with time,
because gas is continuously lost from the galaxy via starburst
outflows (hence reducing $\Sigma_{\rm g}$ in equation~\ref{eq,strip}),
NGC~2276 would become increasingly bulge-dominated with time. The
indication, mentioned in Section~\ref{sec,ram}, that gas in the W disc
is currently being swept inward, suggests that the bulge itself could
be growing at present.  The end product would be a galaxy with
virtually no atomic gas, little star formation, a large bulge-to-disc
ratio, and potentially a more luminous bulge.  These properties are
similar to those of S0 galaxies, suggesting that ram pressure could be
an important mechanism to form such galaxies. We note that the
NGC~2300 group already harbours an S0 galaxy, IC~455 \citep{mulc93},
so whichever mechanisms are at work in the transformation of galaxies
into S0's, they have previously been successful in this environment.
Thus, while there is evidence both for (e.g.\ \citealt{vogt04}) and
against (e.g.  \citealt{goto05}) ram pressure driving the
morphological evolution of cluster spirals into S0's, our results for
NGC~2276 indicate that ram pressure could at least be a contributing
factor, even in environments where it was earlier thought to be
inefficient.

Another class of objects worth mentioning in this context is the
so-called 'E+A' galaxies, believed to be post-starburst galaxies which
have had their starburst activity abruptly truncated within the past
Gyr.  The origin of this rapid halt to their star formation is poorly
understood, but as these galaxies are known to exist also in groups
\citep{blak04}, they could have had their recent evolution affected by
hot intragroup gas.  However, our estimated time-scale for stripping
via IGM interactions in groups of several Gyr suggests that this
process is too slow to provide an important route to forming such
objects.  This would be consistent with recent results attributing the
origin of E+A galaxies predominantly to galaxy-galaxy interactions and
mergers \citep{yama05}.

\section{Conclusions}

A 45~ks {\em Chandra} exposure of the spiral NGC~2276 in the NGC~2300
group of galaxies has revealed a shock-like feature along the western
edge of the galaxy and a low--surface-brightness tail extending in the
opposite direction. The X-ray morphology of NGC~2276 resembles that
seen at optical and radio wavelengths and is suggestive of supersonic
motion of the galaxy through the ambient hot intragroup gas in a
direction towards the west. X-ray imaging spectroscopy reveals
evidence of a build-up of thermal pressure at the leading western edge
of the galaxy, and the density and temperature increase of this gas is
consistent with the presence of a shock front due to the galaxy moving
at Mach~1.7 ($\sim 850$~km~s$^{-1}$) through the surrounding medium.
Diffuse X-ray emission with $kT\approx 0.5$~keV is detected across the
optical disc, and a number of X-ray bright point sources are seen
along the W edge, where strong star formation activity is taking
place. Contrary to previous claims, we conclude that tidal
interactions are not essential to explaining this star formation
activity, nor the morphology of the galaxy as seen across a range of
wavelengths.

In order to investigate the effects of ram pressure resulting from the
motion of NGC~2276 through the intragroup gas, we have modelled the
gravitational potential of the galaxy, and shown that the radius
outside which gas can be stripped by ram pressure corresponds well to
the observed radius of the gaseous and stellar disc. Ram pressure
stripping could therefore have dictated the present size of the disc.
Whereas ram pressure itself is probably insufficient to remove
significant amounts of cold disc gas from the galaxy, turbulent
viscous stripping could be removing $\sim 5$~M$_\odot$~yr$^{-1}$. Due
to the high star formation rate of the galaxy, starburst outflows are
very likely taking place as well, although we cannot confirm this
directly from the X-ray data as the galaxy is viewed nearly face-on.
We estimate that such outflows could be responsible for an additional
mass loss of no more than $\sim 5$~M$_\odot$~yr$^{-1}$, and that these
outflows would also be strongly affected by ram pressure. As the X-ray
tail extending to the east of the galaxy would suggest a mass loss
rate of 3--6~M$_\odot$~yr$^{-1}$, this implies that either of these
two mechanisms could explain the presence of this tail. A comparison
to existing H{\sc i} data indicates that cold H{\sc i} gas is also
being removed from the disc, but that most of the gas stripped from
the galaxy is currently in a hot phase.

At the present mass loss rate, the current H{\sc i} supply in the disc
of NGC~2276 would be exhausted within 1--2 Gyr. From the present H{\sc
  i} content, we further estimate that gas stripping could have been
active for the previous $\sim 2$~Gyr. Hence on reasonably short
time-scales -- a small fraction of a Hubble time -- all H{\sc i} in
this fairly massive spiral could be lost.  This strongly suggests that
the removal of galactic gas through interactions with the surrounding
hot medium {\em can} be effective in group environments, although, in
this particular case, the process is expected to be slower than in
cluster cores by two orders of magnitudes.  A secondary effect of the
ram pressure experienced by NGC~2276 is that it will inhibit the
cooling out of any hot coronal gas on to the disc, thus preventing the
infall of fresh material for continuous star formation.  Combined with
the gas loss, this implies that NGC~2276 would eventually evolve into
an object with properties not too dissimilar from present-day S0
galaxies. Our results therefore suggest that ram pressure could at
least be a contributing factor in the morphological evolution of
spirals into S0's, even in environments where this mechanism was
earlier thought to be inefficient.

\section*{Acknowledgments}
We thank the referee for constructive and insightful comments.  This
work has made use of the NASA/IPAC Extragalactic Database (NED) and
the HyperLeda database. JR acknowledges the support of the European
Community through a Marie Curie Intra-European Fellowship under
contract no.\ MEIF-CT-2005-011171. JSM acknowledges partial support
from Chandra grant G04-5144X and NASA grant NNG04GC846.

\bsp

\label{lastpage}

\end{document}